%% file: paper20261118.tex
\documentclass[prd,aps,twocolumn,a4paper,nofootinbib,longbibliography]{revtex4-1}

\usepackage{graphicx}
\usepackage{amsmath,amsfonts,amssymb}
\usepackage{enumitem}
\usepackage{cleveref}
\usepackage[nohyperlinks]{acronym}

\newlist{filterlist}{enumerate}{1}
\setlist[filterlist]{label=\arabic*.}
\crefname{filterlisti}{filter}{filters}
\Crefname{filterlisti}{Filter}{Filters}

\def\Msun{{\rm M_{\odot}}}
\def\Mmax{{M_{\rm max}^{\rm TOV}}}
\def\Rmax{{R_{\rm max}^{\rm TOV}}}
\def\nmax{{n_{\rm max}^{\rm TOV}}}
\def\veps{\varepsilon}

\def\sro{\texttt{SROEOS}}

\newcommand{\csqp}[1]{c^2_{{\rm PNM},#1}}
\newcommand{\csqs}[1]{c^2_{{\rm SNM},#1}}

\newacro{xeft}[$\chi$EFT]{chiral effective field theory}
\newacro{pQCD}[pQCD]{perturbative quantum chromodynamics}
\newacro{EOS}{equation of state}
\newacroplural{EOS}[EOSs]{equations of state}
\newacro{NS}{neutron star}
\newacro{TOV}{Tolman--Oppenheimer--Volkoff}
\newacro{PNM}{pure neutron matter}
\newacro{SNM}{symmetric nuclear matter}
\newacro{UG}{unitary gas}
\newacro{GP}{Gaussian process}
\newacro{NSE}{nuclear statistical equilibrium}
\newacro{SNA}{single-nucleus approximation}
\newacro{ESS}{effective sample size}
\newacro{KDE}{kernel-density estimate}

\makeatletter
\AtBeginDocument
 {
   \def\ltx@label#1{\cref@label{#1}}%
   \def\label@in@display@noarg#1{\cref@old@label@in@display{#1}}%
\def\label@in@mmeasure@noarg#1{%
    \begingroup%
      \measuring@false%
      \cref@old@label@in@display{#1}%
    \endgroup}%
 } %
\makeatother

\begin{document}

\title{From Multimessenger Inference to Simulations: A Ranked Ensemble of Finite-Temperature Equations of State}

\author{Maximilian \surname{Jacobi}}
\email[Contact author: ]{maximilian.jacobi@uni-jena.de}
\author{Giulia \surname{Huez}}
\author{Sebastiano \surname{Bernuzzi}}
\affiliation{Theoretisch-Physikalisches Institut, Friedrich-Schiller-Universit{\"a}t Jena, 07743, Jena, Germany}
\author{David \surname{Radice}}
\affiliation{Institute for Gravitation and the Cosmos, The Pennsylvania State University, University Park, PA 16802, USA}
\affiliation{Department of Physics, The Pennsylvania State University, University Park, PA 16802, USA}
\affiliation{Department of Astronomy \& Astrophysics, The Pennsylvania State University, University Park, PA 16802, USA}

\date{\today}

\begin{abstract}
We construct a set of microphysical, finite-temperature \acp{EOS} for numerical simulations of neutron star mergers and core-collapse supernovae that is consistent with modern constraints from nuclear theory and multimessenger astronomy and systematically spans the posterior distribution of allowed \acp{EOS}.
The \acp{EOS} are based on a simplified Skyrme functional whose inputs are nuclear matter saturation properties, extended to supra-saturation densities through the speed of sound at a set of reference densities.
The models are assigned continuous likelihood weights from \acl{xeft} and \acl{pQCD} calculations, the gravitational-wave signal GW170817, and four NICER mass--radius measurements.
From the resulting catalog of 3.6 million \acp{EOS}, the posterior yields $R_{1.4} = 11.8^{+0.8}_{-0.7}$\,km, $\Lambda_{1.4} = 342^{+181}_{-118}$, and $\Mmax = 2.17^{+0.23}_{-0.16}\,\Msun$ (medians with $90\%$ credible intervals).
From this posterior we select a ranked twelve-member ensemble, headed by a fiducial, central \ac{EOS}, whose members are individually plausible while bracketing the posterior spread of each neutron-star observable.
For all ensemble members we generate general-purpose finite-temperature tables with the \sro{} code, each accompanied by nucleon effective-mass variants for systematic studies of thermal effects, to be made publicly available upon publication.
\end{abstract}

\maketitle
\acresetall

\section{Introduction}
\label{sec:intro}

Numerical-relativity simulations carry the predictive burden for \ac{NS} mergers \citep{Baiotti:2016qnr,Shibata:2019wef,Radice:2020ddv,Bernuzzi:2020tgt}: the tidal phase of the inspiral waveform is calibrated against them \citep{Bernuzzi:2014owa,Abac:2023ujg}, and the postmerger gravitational-wave spectrum \citep{Breschi:2019srl}, the fate of the remnant \citep{Kashyap:2021wzs,Perego:2021mkd}, and the mass and composition of the ejecta that undergo $r$-process nucleosynthesis and power kilonovae \citep{Nedora:2020hxc,Kiuchi:2022nin,Neuweiler:2025klw,Magistrelli:2025xja} are direct simulation products.
Among the physics inputs to these simulations, none is more consequential than the \ac{EOS} of dense matter, which is highly uncertain in the very density range the mergers probe.
The \ac{EOS} dependence is sharpest in the fate of the merger remnant: whether the remnant collapses promptly, survives briefly, or persists is decided almost exclusively by the \ac{EOS} once the component masses are fixed \citep{Bauswein:2013jpa,Radice:2020ddv}.
That outcome gates every observable after the inspiral: the postmerger gravitational-wave signal, the ejecta and the associated kilonova, and the conditions for launching a jet \citep{Hayashi:2021oxy,Kiuchi:2023obe,Musolino:2024sju}.

Meanwhile, the cold \ac{EOS} itself is increasingly well constrained.
\Ac{xeft} predicts the energy and pressure of neutron matter close to the saturation density with quantified truncation errors \citep{Hebeler:2013nza,Lynn:2015jua,Drischler:2020hwi,Lonardoni:2019ypg,Keller:2022crb,Gottling:2025ohe}, and \ac{pQCD} bounds the pressure at asymptotically high densities, reaching down into the \ac{NS} regime through causality and thermodynamic consistency \citep{Komoltsev:2021jzg,Gorda:2022jvk}.
Astrophysics constrains the region in between, namely via the tidal polarizability determined from GW170817 \citep{LIGOScientific:2017vwq,LIGOScientific:2018cki,LIGOScientific:2018hze}, the NICER mass--radius measurements of four millisecond pulsars \citep{Riley:2019yda,Miller:2019cac,Riley:2021pdl,Miller:2021qha,Salmi:2024aum,Choudhury:2024xbk,Kini:2026rjx,Mauviard:2025dmd}, and the observation of radio pulsars with masses around $2\,\Msun$ \citep{Fonseca:2021wxt}.
Bayesian frameworks combine these data on flexible cold-\ac{EOS} representations (piecewise polytropes \citep{Read:2008iy}, speed-of-sound parameterizations \citep{Tews:2018kmu,Greif:2018njt}, nonparametric \ac{GP} priors \citep{Landry:2018prl}) into posterior distributions over the pressure--density relation and the derived \ac{NS} observables \citep{Annala:2017llu,Legred:2021hdx,Dietrich:2020efo,Huth:2021bsp,Koehn:2024set,Rutherford:2024srk,Mendes:2026mgc}.
A parallel line of work runs the same inference on microphysical nucleonic models, Skyrme-type energy-density functionals \citep{Lim:2018bkq,Lim:2023dbk,Beznogov:2023jqp,Beznogov:2024vcv,Klausner:2024jgu,Klausner:2025ucq,Klausner:2026foh} and the nucleonic metamodel \citep{Margueron:2017eqc,Margueron:2017lup}, tying the posterior directly to nuclear-matter parameters.

Merger simulations have not kept pace with this narrowing.
A typical campaign adopts a handful of the publicly available general-purpose tables \citep[e.g.][]{Radice:2018pdn,Kiuchi:2019kzt,Gonzalez:2022mgo}, drawn from an ecosystem of more than a hundred \citep{Oertel:2016bki,Typel:2013rza} and chosen largely by availability and numerical robustness.
Many of these tables predate the constraints above, and some are in tension with them.
The resulting sets are also heterogeneous (different functional forms, different treatments of inhomogeneous matter, different vintages of nuclear input), so the spread of simulation outcomes across a set confounds \ac{EOS} physics with model heterogeneity, and measures neither a credible interval nor any other defined statistic of the \ac{EOS} uncertainty.
There is, at present, no systematic route from the inferred \ac{EOS} posterior to the \ac{EOS} input of a simulation campaign.

Two obstacles keep the inference and simulation communities working from different \ac{EOS} sets.
The first is how much of the \ac{EOS} each community needs.
The astrophysical observables entering multimessenger inference are properties of cold, isolated \acp{NS}, fixed by the one-dimensional, cold, $\beta$-equilibrated pressure--density relation alone.
Since sampling a posterior moreover takes millions of likelihood evaluations, inference has every reason to parameterize that curve directly rather than model the full composition and temperature dependence it never evaluates.
A simulation instead evolves matter across wide ranges of density, proton fraction, and temperature.
Its complete \ac{EOS} input is a thermodynamically consistent table in all three variables, joined continuously onto a microphysical description of inhomogeneous low-density matter in nuclear statistical equilibrium \citep{Oertel:2016bki}.
This completeness matters most for the ejecta, their nucleosynthesis, and kilonova predictions \citep{Foucart:2016vxd,Foucart:2022bth}, while for the gravitational waveform alone simplified treatments are often adequate.
Promoting an inferred cold curve to such a table after the fact requires an ad hoc thermal closure, usually a constant-$\Gamma_{\rm th}$ prescription, whose error concentrates in exactly the postmerger observables the simulations exist to predict \citep{Bauswein:2010dn,Figura:2020fkj,Fields:2023bhs}.
The $M^*$ framework of \citet{Raithel:2019gws} provides a more consistent closure, modeling the density dependence of the thermal pressure through the nucleon effective mass \citep{Most:2021ktk}, but it still supplements the cold \ac{EOS} with an assumed thermal sector rather than deriving both from a single microphysical model, and, being built on uniform-matter thermodynamics, it does not supply the inhomogeneous low-density matter that a general-purpose table describes.
The second obstacle is cost.
Due to the large computational cost of merger simulations, a campaign fields around ten \acp{EOS} at most, so a posterior catalog of $\mathcal{O}(10^6)$ \acp{EOS} is unusable without a principled way to reduce it to a few members.
This work removes both obstacles with a single pipeline to create a small set of twelve \acp{EOS} that systematically explore the space of \acp{EOS} allowed under current nuclear and astrophysical constraints.
New GPU-accelerated, exascale-capable simulation codes now coming online for mergers and core-collapse supernovae \citep{Fields:2024pob,Shankar:2022ful,Kalinani:2024rbk,Musolino:2026xms,Endeve:2026mmj,Han:2026vnm,Kuroda:2026obe,Wang:2026ytk} open up the possibility of large-scale simulation campaigns, making the need for a systematic set of \acp{EOS} greater than ever.

The existing work closest to ours comes from Beznogov and Raduta, who inferred Bayesian posteriors over extended-Skyrme parameter space \citep{Beznogov:2023jqp,Beznogov:2024vcv} and released families of constraint-compliant finite-temperature tables \citep{Raduta:2025yst,Raduta:2025fpr}, and from \citet{Du:2021rhq}, who propagate parameter probability distributions through a finite-temperature free-energy model and release seven representative tables.
Both programs deliver simulation-ready tables, but neither connects them to a modern multimessenger inference, as neither includes the constraints from GW170817 and NICER.
The Beznogov--Raduta constraints all act at or below saturation density, with astrophysics entering only as a maximum-mass cut, and the high-density input of \citet{Du:2021rhq} is frozen to X-ray measurements predating these missions.
Other programs fix a small set of models deterministically rather than sampling a posterior.
\citet{Alford:2022bpp,Alford:2023rgp}, for example, fit relativistic mean-field models to \ac{xeft} neutron matter and to saturation properties and release four finite-temperature tables that span a range of stiffness and are consistent with astrophysical constraints, without quantifying how much of the allowed \ac{EOS} space they cover.
\citet{Li:2023bid} vary the symmetry-energy slope and the skewness on a grid around three covariant density functionals and release the resulting families of cold $\beta$-equilibrium tables, which a simulation has to supplement with a thermal prescription.
\citet{Jost:2026wcd} build finite-temperature tables on the functional of \citet{Huth:2020ozf} and vary the saturation parameters, the high-density pressure and the effective mass one at a time around a single baseline, checking each variation against the pressure band of a multimessenger inference \citep{Rutherford:2024srk}.

We run forward-model Bayesian inference directly on the Skyrme-type functional of the \sro{} code \citep{Schneider:2017tfi}, so every \ac{EOS} we construct can be propagated by \sro{} into a complete general-purpose finite-temperature table.
Traditional Skyrme functionals typically struggle to be flexible enough at supranuclear densities and overestimate the correlation of the high-density \ac{EOS} with saturation-point properties.
Furthermore, they usually adopt a small, fixed set of density exponents \citep{Schneider:2017tfi,Lim:2018bkq}, which fail to produce a maximum in the speed of sound and instead tend to turn superluminal at densities reached in \ac{NS} cores \citep{Duan:2023amg,Beznogov:2024vcv}.
Such a maximum is, however, favored by \ac{pQCD}, whose constraints propagate down to \ac{NS} densities and soften the \ac{EOS} there \citep{Komoltsev:2021jzg,Gorda:2022jvk}, although the size of this effect depends on the density at which the \ac{pQCD} input is imposed \citep{Somasundaram:2022ztm}.
We therefore define the prior by uniformly drawing six nuclear saturation properties and the speed of sound at a set of supra-saturation reference densities, which together determine the functional's coefficients.
Prescribing the speed of sound directly keeps the supra-saturation sector flexible and essentially uncorrelated with the saturation properties.
For every draw we redetermine the exponents to ensure that the prescribed quantities are realized by a thermodynamically consistent \ac{EOS} that stays causal up to ten times the saturation density.
Compared with fixed exponents, this widens the prior to a much larger range of high-density behavior, so that the posterior at the densities mergers probe is shaped by the astrophysical constraints and less by the choice of parameterization.
Each \ac{EOS} in the resulting catalog is then reweighted by a product of likelihoods from nuclear physics and astrophysics, yielding the \ac{EOS} posterior on which the rest of the analysis builds.
From this weighted catalog we extract a twelve-member ensemble by a nested max--min construction.
Each member is additionally delivered with a set of nucleon effective-mass variants that vary the thermal response at fixed cold-sector behavior.

Throughout the sampling and reweighting, each \ac{EOS} is represented only by a cheap one-dimensional cold $\beta$-equilibrium table with a simply matched crust.
For the twelve selected members, we use the \sro{} code to propagate the same Skyrme coefficients into complete three-dimensional tables in density, temperature, and composition, deriving the thermal sector and the inhomogeneous low-density matter from the same functional.
In these tables the crust is described by the \ac{SNA} and its low-density part is determined by a \ac{NSE} ensemble of thousands of nuclides with measured masses.
These general-purpose tables will be made publicly available on Zenodo upon publication of this article.
Although this work mostly targets \ac{NS} mergers, the tables are not merger-specific.
Core-collapse supernova simulations require the same general-purpose finite-temperature \ac{EOS} tables spanning wide ranges of density, composition, and temperature \citep{Oertel:2016bki}, and shock revival, protoneutron-star contraction, and black-hole formation respond to the same uncertain high-density physics \citep{Yasin:2018ckc,Schneider:2019shi}.

The paper is organized as follows.
\Cref{sec:model} specifies the functional, the prior on its parameters, and the exponent existence search to obtain a valid \ac{EOS} for each draw, and \cref{sec:likelihoods} describes the likelihood reweighting.
\Cref{sec:results} presents the production catalog, the posterior, and its robustness, and \cref{sec:selection} constructs the simulation ensemble and quantifies how well it brackets the posterior.
\Cref{sec:conclusions} summarizes the results, compares our work with constraint-compliant table sets from the literature, and discusses the limitations of the approach and its natural extensions.

\section{EOS model and priors}
\label{sec:model}

All \acp{EOS} in this work are generated with the \sro{} code \citep{Schneider:2017tfi}, which turns a Skyrme-type parameterization into a complete general-purpose finite-temperature table.
To our knowledge, it is the only publicly available code that does so for an arbitrary parameterization, with a self-consistent thermal sector and a microphysical treatment of inhomogeneous matter at low densities.
It also places no limit on the number of density-dependent terms in the functional, which we exploit to make the supra-saturation \ac{EOS} far more flexible than standard few-term Skyrme parameterizations allow.

The underlying functional gives the internal energy density $e$ as
\begin{align}
  e (n, x, T) =& \sum_t \frac{\tau_t(n, x, T)}{2 m_t^*(n)} - x n \Delta \nonumber\\
                 &+ \sum_i \left[a_i + 4x(1-x) b_i\right] n^{\delta_i +1}\,,
  \label{eq:skyrme_funct}
\end{align}
where $n$ is the baryon number density, $x = n_p/n$ is the proton fraction ($x \equiv Y_e$ in the astrophysical context), $T$ is the temperature, $t \in \{p,n\}$ labels the nucleon species, $\tau_t$ is the kinetic energy density of species $t$, $\Delta = m_n - m_p$ is the neutron-proton mass difference, and $a_i$, $b_i$, $\delta_i$ are free parameters.
Note that we have absorbed the coefficients $a, b, c_i, d_i$ from \citet{Schneider:2017tfi} into common $a_i$ and $b_i$, with the result that the first exponent is fixed at $\delta_1 = 1$.

The Landau effective mass $m_t^*$ is
\begin{align}
  \label{eq:meff}
  \frac{1}{2 m_t^*(n)} = \frac{1}{2 m_t} + \alpha_1 n_t + \alpha_2 n_{-t}\,,
\end{align}
where $m_t$ is the vacuum mass of species $t$, $n_t$ its number density ($n_p = xn$, $n_n = (1-x)n$, and $n_{-t}$ denotes the density of the opposite species), and $\alpha_1$, $\alpha_2$ are additional free parameters.
None of the data employed in this work directly constrain $m^*$, and the effective mass acts mainly on the thermal sector, characterized by the thermal index
\begin{align}
  \label{eq:gammath}
  \Gamma_{\rm th}(n,x,T) = 1 + \frac{P(n,x,T) - P(n,x,0)}{e(n,x,T) - e(n,x,0)}\,,
\end{align}
with $P$ and $e$ the total pressure and energy density.
The effective mass's imprint on the cold \ac{EOS} through $P_{\rm kin}$ is small and easily compensated by the interaction terms.
Rather than sampling parameters our data cannot constrain, we fix the effective masses at saturation density to sector-specific reference values from the literature: the neutron effective mass in \ac{PNM} ($n = n_0$, $x = 0$) to $m^*_n/m_n = 0.95$, the center of the chiral many-body-perturbation-theory band \citep{Keller:2020qhx,Carbone:2019pkr}, and the effective mass in \ac{SNM} ($x = 1/2$) to $m^*/m = 0.85$, within the isoscalar range $0.80 \pm 0.10$ spanned by many-body calculations and optical-potential analyses \citep{Li:2018lpy,Carbone:2019pkr}.
These two conditions fix $\alpha_1$ and $\alpha_1 + \alpha_2$ (hence $\alpha_2$) in closed form for every draw, and imply $m^*_p < m^*_n$ in neutron-rich matter, the correct sign of the isospin splitting \citep{Li:2018lpy}.
The released simulation ensemble additionally carries effective-mass variants that shift both sector targets by $\pm 0.10$ for each \ac{EOS} at fixed cold-sector behavior (\cref{sec:selection}).
The monotonic density profile of \cref{eq:meff} is, however, too restrictive to reproduce finite-temperature \ac{xeft} calculations of neutron matter \citep{Keller:2020qhx,Keller:2022crb}, in which three-nucleon forces make the effective mass rise again above about $0.1\,{\rm fm^{-3}}$.
Matching the effective mass at saturation therefore does not make the thermal response consistent with \ac{xeft}, and the thermal sector of our tables lies outside these constraints.
An \ac{xeft}-consistent thermal sector requires a microscopically motivated effective mass, as in the functional of \citet{Huth:2020ozf} and the tables built on it by \citet{Jost:2026wcd}.

Henceforth, we consider only cold ($T = 0$) nuclear matter.
The cold pressure is given by
\begin{align}
  \label{eq:pressure}
  P (n, x) =& \; \left. n^2 \frac{\partial \veps}{\partial n} \right|_x \nonumber\\
  =& \; P_{\rm kin}(n,x) + \sum_i \left[a_i + 4x(1-x) b_i\right] \delta_i\, n^{\delta_i+1}\,,
\end{align}
where $\veps = e/n$ is the energy per baryon and the related kinetic pressure is
\begin{align}
  \label{eq:pkin}
  P_{\rm kin}(n,x) = \sum_t \tau_t \left[\frac{1}{3 m_t^*} + \alpha_1 n_t + \alpha_2 n_{-t}\right]\,,
\end{align}
with the cold kinetic energy density
\begin{align}
  \label{eq:tau_0}
  \tau_t = \frac{3}{5} (3\pi^2 n_t)^{2/3} n_t\,.
\end{align}

Our Skyrme functional has sixteen free linear coefficients ($a_i, b_i$) and seven free exponents ($\delta_i$).
We reparameterize the functional in terms of sixteen prescribed physical quantities given in \cref{tab:ranges}, namely six saturation parameters, described in \cref{sec:prior:saturation}, and ten speed-of-sound targets at five supra-saturation anchor densities, described in \cref{sec:prior:cs2}, from which the sixteen linear coefficients are recovered by solving a linear system.
Given the set of exponents, whose determination is described in \cref{sec:search}, the chosen physical parameters uniquely determine the linear coefficients via two decoupled linear systems, one per isospin channel, given in \cref{app:linear_system}.
\begin{table}
  \centering
  \caption{\label{tab:ranges}
    Sampling ranges for the prior.
    All parameters are drawn uniformly and independently.
    $n_0$ denotes absolute baryon density in ${\rm fm}^{-3}$.}
  \begin{tabular}{lcc}
    Parameter & min & max \\
    \hline
    $n_0$ [fm$^{-3}$] & 0.155 & 0.165 \\
    $B$ [MeV]   & $-$16.5 & $-$15.5 \\
    $K$ [MeV]   & 160 & 315 \\
    $J$ [MeV]   & 28.6 & 38.1 \\
    $L$ [MeV]   & 20 & 90 \\
    $\tilde{K}_{\rm sym}$ [MeV] & $-$400 & 800 \\
    \hline
    $c_s^2$ at $2,4,6,8,10\,n_0$ & 0 & 1 \\
  \end{tabular}
\end{table}

\subsection{Saturation parameters}
\label{sec:prior:saturation}

The six saturation parameters of \cref{tab:ranges} are the saturation density $n_0$, the binding energy $B$ and the incompressibility $K$ of \ac{SNM}, the symmetry energy $J$, its slope $L$, and a sixth parameter that sets the stiffness of neutron matter at saturation.
For the latter we sample the difference between the incompressibilities of \ac{PNM} and \ac{SNM},
\begin{equation}
  \label{eq:ktilde_sym}
  \tilde{K}_{\rm sym} = 9\left.\frac{\partial P}{\partial n}\right|_{n_0,\,x=0} - K\,.
\end{equation}
We write $\tilde{K}_{\rm sym}$ to avoid confusion with the symmetry-energy curvature commonly quoted in the literature, $K_{\rm sym} = 9 n_0^2\, \mathrm{d}^2 E_{\rm sym}/\mathrm{d}n^2|_{n_0}$ with $E_{\rm sym}(n) = \veps(n, 0) - \veps(n, 1/2)$, for which we reserve the symbol $K_{\rm sym}$.
The two differ by an offset proportional to the slope, $\tilde{K}_{\rm sym} \simeq K_{\rm sym} + 6L$, which ranges from $122$ to $542$\,MeV across the catalog (median $+345$\,MeV).
We sample $\tilde{K}_{\rm sym}$ and report results in terms of $K_{\rm sym}$.

Two considerations set the prior range widths in \cref{tab:ranges}.
First, for several saturation parameters ($J$, $L$, $K$, $\tilde{K}_{\rm sym}$) determinations from different probes disagree well beyond their quoted uncertainties, so no single measurement can serve as a prior.
Second, the data we employ in \cref{sec:likelihoods} constrain these parameters very unevenly: $K$ is left essentially unconstrained (its reweighted marginal reproduces the prior), while $J$ is narrowed by the chiral energy term to $30.7$--$35.4$\,MeV.
However, $L$ and $\tilde{K}_{\rm sym}$, which control the pressure of neutron-rich matter near saturation, are the parameters one would naturally expect to imprint most strongly on \ac{NS} observables.
Furthermore, they are directly constrained by the \ac{xeft} band employed in \cref{sec:nuclear_likelihoods}.
We therefore adopt deliberately broad uniform sampling intervals (``boxes'' hereafter) and let the likelihoods, not the box edges, carry the constraining power.

The saturation density $n_0$ (defined by $P(n_0, 1/2) = 0$) and the binding energy $B = \veps(n_0, 1/2)$ fix two SNM degrees of freedom.
Both are well constrained by nuclear masses and charge-radius systematics: $n_0 \approx 0.160 \pm 0.005\,{\rm fm}^{-3}$ and $B \approx -16.0 \pm 0.5\,{\rm MeV}$ \citep{Li:2019xxz}.
Our boxes match these quoted uncertainties exactly.
The incompressibility $K = 9\,\partial P/\partial n|_{n_0, 1/2}$ is traditionally determined from isoscalar giant monopole resonances, with a long-standing consensus value of $K \approx 240 \pm 20$\,MeV \citep{Garg:2018uam}.
Determinations beyond that canon disagree well outside its error bar.
A recent reanalysis of the monopole data, the first to describe the tin and lead isotopes consistently, favors the lower $K \approx 227$\,MeV \citep{Li:2022suc}.
Heavy-ion flow measurements prefer $K = 190 \pm 30$\,MeV \citep{LeFevre:2015paj,Huth:2021bsp}, \ac{xeft} predicts $K = 260 \pm 54$\,MeV \citep{Drischler:2020yad}, Bayesian Skyrme surveys report posteriors near $260$\,MeV \citep{Beznogov:2024vcv}, and an analysis of the surface-term systematics of the monopole extraction argues for values as high as $250$--$315$\,MeV \citep{Stone:2014wza}.
Our range $160$--$315$\,MeV is the envelope of the quoted $\pm1\sigma$ intervals of these determinations: the lower edge is the reach of the heavy-ion constraint ($190 \pm 30$\,MeV), the upper edge the top of the surface-systematics range of \citet{Stone:2014wza}, with the \ac{xeft} interval (reaching $314$\,MeV) sitting just inside.
The wide envelope does not hinder the precision of our inference.
The data we employ leave $K$ unconstrained (the reweighted marginal reproduces the prior), and the inferred observables are insensitive to the range (see \cref{sec:results:robustness}).

The symmetry energy $J = \veps(n_0, 0) - B$, its slope $L = 3P(n_0, 0)/n_0$, and $\tilde{K}_{\rm sym}$ fix the three \ac{PNM} degrees of freedom.

The symmetry energy is among the better-constrained saturation observables: nuclear structure systematics and \ac{xeft} calculations broadly favor $J \approx 30$--$35$\,MeV \citep{Drischler:2021kxf,Li:2019xxz}, while the PREX-2 neutron-skin measurement of $^{208}$Pb yielded the notably larger value $J = 38.1 \pm 4.7$\,MeV \citep{Reed:2021nqk}, and analyses dominated by the CREX $^{48}$Ca skin measurement \citep{CREX:2022kgg} fall slightly below the consensus, e.g.\ $J = 29.1^{+2.1}_{-1.8}$\,MeV \citep{Zhang:2022bni}.
Within any interval of this size the astrophysical likelihoods of \cref{sec:likelihoods} leave $J$ unconstrained, and only the chiral energy term of \cref{sec:nuclear_likelihoods} narrows it.
We tie both box edges to literature values. The lower edge is the unitary-gas floor $J \ge 28.6$\,MeV \citep{Tews:2016jhi}, a hard theoretical floor, and the upper edge is the PREX-2 central value $38.1$\,MeV, the largest central value any modern determination supports.
The reweighted marginal lies more than $2$\,MeV inside both edges, so neither shapes the posterior.
Here, the envelope runs up to the PREX-2 central value only (not over the $\pm1\sigma$ intervals as for $K$), because PREX-2 is an outlier among $J$ determinations, and its $+1\sigma$ reach ($42.8$\,MeV) would let a single tension-laden measurement stretch the box $\sim$5\,MeV beyond every other determination's reach (the next-highest $+1\sigma$ endpoint lies near $35$\,MeV).
The floor derives from the unitary-gas bound \citep{Tews:2016jhi},
\begin{equation}
  \label{eq:ug_bound}
  \veps_{\rm PNM}(n) \ge \xi\,\veps_{\rm FG}(n)\,,
\end{equation}
where $\veps_{\rm FG}$ is the free Fermi-gas energy per baryon and $\xi = 0.376$ the Bertsch parameter.
The constraint states that the energy per particle of cold dilute neutron matter cannot drop below that of the unitary Fermi gas.
The same bound re-appears as a hard constraint of the existence search (\cref{sec:search}), enforced there as a function of density rather than only through $J$.

The slope $L$ is considerably less tightly constrained: global analyses of nuclear observables find $L \approx 40$--$80\,{\rm MeV}$ with substantial scatter \citep{Li:2019xxz,Lattimer:2012xj}, and PREX-2 implies $L \approx 106 \pm 37\,{\rm MeV}$ \citep{Reed:2021nqk}, in tension with most other probes.
The CREX $^{48}$Ca measurement pulls in the opposite direction, favoring $L \lesssim 40$\,MeV \citep{CREX:2022kgg,Zhang:2022bni}, and the two skins have so far resisted a joint description within a single density functional \citep{Reinhard:2022inh}.
Analyses that combine the skin data with other probes, or that reinterpret the PREX-2 asymmetry together with the $^{208}$Pb dipole polarizability, converge on $L \approx 50$--$60$\,MeV \citep{Reinhard:2021utv,Essick:2021kjb,Lattimer:2023rpe}, as does \ac{xeft} ($L = 59.8 \pm 4.1$\,MeV, \citealp{Drischler:2020yad}).
We therefore choose a very broad box for $L$, $20$--$90$\,MeV, and rely on the \ac{xeft} pressure band (\cref{sec:nuclear_likelihoods}) to shape the marginal, as the energy term does for $J$.
In the production catalog no hard filter imposes a floor on $L$. Survivors populate the box down to its $20$\,MeV edge, with about $0.7\%$ of the catalog below $23$\,MeV, and the likelihood reweighting alone closes the tail, leaving the posterior-weighted 5th percentile at $L = 31.7$\,MeV and only $0.26\%$ of the posterior mass below $23$\,MeV.

Finally, we prescribe $\tilde{K}_{\rm sym}$.
No determination of it is precise enough to motivate box edges, so we again take a deliberately broad box, $[-400, 800]$\,MeV, and rely on the unitary-gas bound (\cref{eq:ug_bound}) and the \ac{xeft} likelihood (\cref{sec:nuclear_likelihoods}) to shape the final posterior.

\subsection{Speed-of-sound anchors}
\label{sec:prior:cs2}

At five reference densities $n_{\rm ref} \in \{2,4,6,8,10\}\,n_0$ we prescribe the squared speed of sound $c_s^2$ of cold nuclear matter at fixed proton fraction, including electrons.
At each anchor density we sample the target $c_s^2(n_{\rm ref}, x=0)$ for \ac{PNM} directly, and independently $c_s^2(n_{\rm ref}, x=1/2)$ for \ac{SNM}.
The five anchors are evenly spaced in units of $n_0$, extending to $10\,n_0$ so that causality is enforced well past the densities any surviving star's core actually reaches (\cref{sec:search}).
All ten targets are drawn uniformly and independently from the full causal interval $c_s^2 \in [0, 1]$, with no ordering imposed between anchors ($c_s^2(n)$ does not need to be monotonic).
In \cref{sec:results:robustness} we test how much this modeling assumption of the prior influences the posterior by reweighting the prior to obtain a measure that is uniform in the pressure instead.

A natural cut-off for the density range over which the \acp{EOS} must be valid is the central density $\nmax$ of the maximum-mass \ac{TOV} star, which typically lies at $\nmax \approx 4$--$7\,n_0$ (\cref{sec:results:robustness}).
The placement of the two topmost anchors is therefore mostly a precaution, ensuring a physical \ac{EOS} up to black-hole formation in simulations, and does not shape the astrophysically relevant part of the \ac{EOS}.

As a direct test of sensitivity to the anchor scheme itself, a companion catalog using three anchors at $\{2, 5.5, 9\}\,n_0$ (evenly spanning the same posterior-relevant density range with a coarser grid) is generated at approximately the same statistics under the otherwise identical pipeline, and the headline results are compared across the two schemes (\cref{sec:results:robustness}).

Not every combination of saturation parameters and speed-of-sound targets drawn from the prior can be realized.
The functional has a fixed form, and for most combinations no \ac{EOS} of that form meets all targets while staying causal and physical between the anchors.
With fixed exponents the excluded part of the prior volume would be larger still.
We therefore leave the exponents free and search them anew for every draw, which makes as many combinations as possible realizable.

\subsection{Exponent existence search}
\label{sec:search}

A draw consists of the sixteen parameters of \cref{tab:ranges}, the six saturation parameters and the ten speed-of-sound targets.
For any choice of the exponent vector $[\delta_i]$ these determine the coefficients of the functional through a linear system (\cref{app:linear_system}), so the prescribed values are met exactly.
The exponents themselves remain free, and they control how the \ac{EOS} behaves between and beyond the anchors.
For each draw we search the exponent space for a vector under which the \ac{EOS} is causal and physical up to $n = 10\,n_0$, and reject the draw if no such vector can be found.
The search treats the exponents as free variables and minimizes a smooth hinge penalty on causality violations (given explicitly in \cref{eq:search_objective} of \cref{app:search}), i.e.\ on excursions of the squared sound speed
\begin{equation}
  \label{eq:cs2_def}
  c_s^2(n,x) = \frac{(\partial P/\partial n)_x}{h(n,x)}
\end{equation}
outside $[0,1]$, evaluated on a fine grid of densities and compositions, where the derivative in the numerator includes the electron pressure and $h = m_n + \veps + P_{\rm b}/n + x\,\mu_e$ is the specific enthalpy per baryon, including the neutron rest mass and, via charge neutrality ($n_e = nx$), the electron enthalpy $x\,\mu_e$. Here $P_{\rm b}$ is the baryonic pressure and $\mu_e$ the chemical potential of the free ultrarelativistic electron gas.
Three further requirements enter as hard constraints on the optimizer rather than penalty terms.
The unitary-gas bound of \cref{eq:ug_bound} is enforced for $n \in [0.01, 1.5]\,n_0$, and the symmetry energy is required to be positive over the full density range.
The third bounds the low-density magnitude of the density-dependent effective couplings $v_a(n) = \sum_i a_i n^{\delta_i - 1}$ and $v_b(n) = \sum_i b_i n^{\delta_i - 1}$ to $|v_{a,b}| \le 500\,{\rm MeV\,fm^3}$ for $n \in [10^{-4}, 0.05]\,{\rm fm^{-3}}$, and a second optimizer pass polishes every accepted exponent vector toward smaller $|v_{a,b}|$ at fixed causality.
Without this bound the search routinely returns exponents clustered just above one with large, nearly canceling coefficients, which satisfy the sound-speed anchors above $0.5\,n_0$ but leave couplings of $10^3$ to $10^{10}\,{\rm MeV\,fm^3}$ at low density, breaking the two-phase solve of the \sro{} crust below ${\sim}0.05\,{\rm fm^{-3}}$.
The implementation is an outer loop of a local, gradient-based constrained optimizer (Sequential Least-Squares Programming, SLSQP) over the free exponents around the inner linear solve outlined in \cref{app:linear_system}.
The full protocol (objective grids, restart budget, convergence criteria) is given in \cref{app:search}.

A failed search does not prove that no valid exponent vector exists.
Certifying nonexistence would require a global search of the seven-dimensional exponent space for every draw, which is neither feasible nor necessary.
Instead, the restart and polish protocol of \cref{app:search} is sized to the point of diminishing returns, beyond which additional attempts recover practically no further solutions.
A rejected draw is therefore one for which no valid vector was found at this fixed search effort, frozen throughout production.
We find that the distinction matters little in practice, as the boundary between accepted and rejected draws is sharp in every two-dimensional slice of the sampled parameters we inspected, so the search acts as a nearly deterministic feasibility criterion.

Each accepted \ac{EOS} is then solved for its stellar structure.
Cold neutron-star matter is in $\beta$ equilibrium, which fixes the proton fraction $x_\beta(n)$ at each density through
\begin{equation}
  \label{eq:beta_eq}
  \left.\frac{\partial \veps}{\partial x}\right|_{x=x_\beta} + \mu_e(nx_\beta) = 0\,.
\end{equation}
Evaluating the functional along this track and adding the electron contribution gives the pressure and energy density as functions of the baryon density alone.
We truncate this cold \ac{EOS} before any loss of monotonicity or causality and match it to the BPS crust \ac{EOS} \citep{Baym:1971pw} at low density (\cref{app:bps}).
The \ac{TOV} equations \citep{Tolman:1939jz,Oppenheimer:1939ne} are then integrated for a series of 50 central pressures to produce the full $M$--$R$ sequence.
Note that we use the matched BPS crust only for the cold \ac{TOV} solutions employed during the catalog stage.
The finite-temperature tables published alongside this work build the low-density inhomogeneous phase within \sro{} itself, including the match to an \ac{NSE} ensemble (\cref{sec:selection}).
Finally, the \ac{EOS} is required to remain valid (monotonic, causal, positive symmetry energy) up to $1.1 \times \nmax$, the central density of the maximum-mass configuration.
Since the search already certifies causality to $10\,n_0 > \nmax$, this removes almost exclusively exceptionally soft candidates with \mbox{$\Mmax \ll 1.97\,\Msun$}, which are practically excluded by the pulsar mass measurements.
Beyond that, no maximum-mass cut is applied.

A last requirement concerns the inhomogeneous matter at low density, which the catalog stage does not model but which every finite-temperature table built from the functional contains.
The \ac{SNA} treatment of inhomogeneous matter used for the released tables (\cref{sec:selection}) describes nuclei as drops of a dense liquid phase in equilibrium with a dilute gas, and it fails for functionals that sustain a liquid--gas coexistence even in nearly pure neutron matter, where no bound drops exist.
For such an \ac{EOS} the single-nucleus solve finds only very light drops in the inner crust at low proton fraction, rejects them, and falls back to uniform matter, which is mechanically unstable there, so the table develops negative sound speeds around neutron drip.
We therefore reject an \ac{EOS} when its functional has a liquid--gas critical point at a proton fraction of $0.02$ with a critical temperature above $4.0$\,MeV.
This crust-stability criterion depends on the coefficients of the functional alone and is applied to every candidate that meets the requirements above.
A critical point exists at that proton fraction for $12.5\%$ of the candidates, and $10.3\%$ of the candidates lie above the threshold and are removed.
We validated the threshold on $80$ \ac{SNA} crust builds spanning the range of critical temperatures.
The fraction of thermodynamically unstable cells rises steadily with the critical temperature, every build with more than $1\%$ unstable cells ($34$ of $80$) lies above the threshold, and six builds above the threshold have fewer.
An \ac{EOS} that fails the criterion is unlikely to yield a clean finite-temperature table, so we remove it from the catalog at this stage.

A second requirement of the same kind concerns the potential part of the symmetry energy, $-\sum_i b_i n^{\delta_i}$.
It sets the sign of the difference between the neutron and proton single-particle potentials, $U_n - U_p$, which the finite-temperature tables carry and which shifts the charged-current neutrino rates in simulations \citep{Martinez-Pinedo:2012eaj,Roberts:2012um}.
In neutron-rich matter this difference is positive for the interactions in common use, and a functional whose potential symmetry energy turns negative at low density reverses it in the dilute gas of the inner crust.
We therefore require the potential symmetry energy to be positive for $n \in [0.001, 0.05]\,{\rm fm^{-3}}$, the density range of that gas, where the two-phase solve of the tables operates, which removes $17.2\%$ of the candidates that pass the crust-stability criterion.
The posterior of \cref{sec:results} and the simulation ensemble of \cref{sec:selection} are thus built on the same set of \acp{EOS}.

The construction and filter steps outlined above define the prior on \ac{EOS} space implicitly: we sample the sixteen-dimensional box of \cref{tab:ranges} uniformly and keep the sub-volume on which the existence search succeeds and all requirements are met.
This defines a curved but sharp restricted volume in the box.
Because the accepted sub-volume is not a product of the marginal intervals, the marginalized prior distribution of each sampled parameter is not uniform, even before any likelihood is applied.

\section{Constraints as importance weights}
\label{sec:likelihoods}

Every data-driven constraint enters the analysis the same way: as a continuous likelihood evaluated per \ac{EOS} of the prior catalog and applied as an importance-sampling weight \citep[e.g.][ch.~10]{Gelman:2013bda}, never as a hard cut.
Note that this forward Monte-Carlo approach is sufficient for our purpose because the constraints retain substantial overlap with the unweighted catalog distribution, so the reweighted posterior stays well-sampled (\ac{ESS} $\sim 2.9\times10^4$; \cref{sec:results:catalogue}) rather than concentrating in an undersampled tail of the prior.
This section specifies the likelihoods, nuclear theory first, then the astrophysical measurements.

\subsection{Nuclear theory}
\label{sec:nuclear_likelihoods}

For each \ac{EOS} in the catalog, i.e.\ each draw that passed the validity requirements of \cref{sec:search}, we compute a continuous weight from a Gaussian likelihood built on the published N$^3$LO \ac{xeft} band of \citet{Gottling:2025ohe}.
We do not train a \ac{GP} ourselves. We take the published pointwise means and uncertainties and model the density correlations between them with a fixed \ac{GP} kernel, as detailed below.
This is the only point in the pipeline where \ac{xeft} information enters explicitly.

The underlying \ac{GP} of \citet{Gottling:2025ohe} is defined on the energy per particle $E/A(n,x)$ with a squared-exponential (RBF) covariance kernel.
Since pressure is a linear functional of $E/A$, namely $P = n^2\,\partial(E/A)/\partial n$, it inherits Gaussian marginals.
However, the full trained \ac{GP} (the fitted kernel hyperparameters and the joint density--composition covariance) is not part of the public deposit, which provides pointwise means and standard deviations of derived curves.

We evaluate the likelihood on the $\beta$-equilibrium pressure $P_\beta(n)$, taking the tabulated mean $\mu(n)$ and standard deviation $\sigma(n)$ directly from Zenodo~\citep{Gottling:2026zen} and modeling the missing density correlations explicitly with an RBF kernel of assumed length scale.
The log-probability of the pressure curve is
\begin{equation}
  \label{eq:log_L_xft}
  \ln\mathcal{L}_{\rm xEFT} = -\frac{1}{2}\,\mathbf{z}^\top R^{-1}\mathbf{z}
                                - \frac{1}{2}\ln|R|\,,
\end{equation}
where the residual vector has components
\begin{equation}
  z_i = \frac{P_\beta(n_i) - \mu(n_i)}{\sigma(n_i)}
\end{equation}
and $R_{ij}$ is the RBF correlation matrix,
\begin{equation}
  R_{ij} = \exp\!\left(-\tfrac{1}{2}\,\frac{(n_i-n_j)^2}{l^2}\right) + \nu\,\delta_{ij}\,,
\end{equation}
with length scale $l = 0.04$\,fm$^{-3}$ (matching the \ac{xeft} training-grid spacing) and diagonal nugget $\nu = 0.05$.
We test the sensitivity of our results to the assumed value of $l$ in \cref{sec:results:robustness}.
At this length scale the 20 evaluation points are spaced closely enough that adjacent RBF correlations exceed $0.98$, so the pure-RBF $R$ is numerically near-singular.
The nugget both regularizes the inverse and represents the small point-to-point uncorrelated component not captured by the smooth part (tabulation error and departures from an exactly-RBF kernel).
Results are insensitive to its precise value.
The \ac{GP} is evaluated at 20 points uniformly spaced in $n \in [0.58, 1.5]\,n_0$.
The $\ln|R|$ term cancels in relative comparisons and is only included for completeness.

The pressure band constrains the slope of the neutron-matter energy per particle, since $P = n^2\,\partial(E/A)/\partial n$, but not its value, which is the integration constant the pressure cannot see.
We therefore add a second Gaussian term on the neutron-matter energy at $n = 0.10\,{\rm fm^{-3}}$, $\ln\mathcal{L}_E = -(E_{\rm PNM} - \mu_E)^2 / 2\sigma_E^2$, with $\mu_E = 11.07$\,MeV from the N$^3$LO \ac{PNM} band of the same deposit and $\sigma_E = 1.0$\,MeV, wider than the tabulated $0.39$\,MeV to allow for the spread between chiral interactions.

Analogously to the \ac{xeft} reweighting above, we compute a continuous weight for each surviving \ac{EOS} from the marginalized \ac{pQCD} likelihood of \citet{Komoltsev:2023zor}, built on the thermodynamic-consistency construction of \citet{Komoltsev:2021jzg}, evaluated only at the \ac{EOS}'s own maximum-\ac{TOV}-mass central density $\nmax$.

\subsection{Astrophysical likelihoods}
\label{sec:astro}

In addition to the nuclear-theory-derived likelihood weights of \cref{sec:nuclear_likelihoods}, we confront the catalog with astrophysical measurements through a further set of continuous weights, computed directly for every \ac{EOS}.
Five astrophysical inputs enter the baseline analysis, the GW170817 posterior and the NICER mass--radius posteriors of four millisecond pulsars.
Each of them is available as a set of posterior samples.
We turn these into a likelihood by fitting a \ac{KDE} to the samples and evaluating it at the radii or tidal deformabilities that a candidate \ac{EOS} predicts, marginalized over the source masses.
The estimator is the same for all five inputs and is detailed in \cref{app:estimator}.

The GW170817 likelihood is built from a posterior over the component masses and dimensionless tidal polarizability $\Lambda(M) = \tfrac{2}{3}\,k_2\,C^{-5}$, with $k_2$ the quadrupole tidal Love number and $C = GM/(Rc^2)$ the compactness.
We compute it by performing parameter estimation with the \texttt{bajes} code~\citep{Breschi:2021wzr}.
The priors and the settings for the analyses of the GW data~\citep{LIGOScientific:2018mvr} are the same as in~\cite{Breschi:2024qlc} and we employ the effective-one-body model, \texttt{TEOBResumSPA}~\citep{Nagar:2018gnk,Gamba:2020ljo}.
As a sensitivity check (\cref{sec:results:robustness}), we replace our posterior by the GW170817 posterior of the LIGO--Virgo Collaboration \citep{LIGOScientific:2018hze}, obtained with the \texttt{IMRPhenomPv2\_NRTidal} waveform model.
It differs from ours in the waveform model, the sampling code and the spin prior, which unlike ours admits only low spins, and enters the likelihood in the same way.

Four NICER mass--radius posteriors enter: J0740+6620 \citep{Salmi:2024aum}, J0437--4715 \citep{Choudhury:2024xbk}, J0030+0451 with the \texttt{PDT-U} hotspot-model samples of the six-year reanalysis \citep{Kini:2026rjx}, and J0614--3329 \citep{Mauviard:2025dmd}.
For J0437 we adopt the X-PSI analysis used throughout the recent inference literature, although we note that an independent reanalysis that additionally models the modulated nonthermal emission component favors somewhat larger radii \citep{Miller:2025qfq}.
For J0614 we likewise adopt the \texttt{PDT-U} hotspot-model samples, the marginally evidence-preferred solution of \citet{Mauviard:2025dmd}, matching the evidence-led choice for J0030.
An independent analysis of the same NICER data by \citet{Miller:2026vpr} recovers a consistent but appreciably broader radius, a symmetric $68\%$ credible range of $9.88$--$12.77$\,km against $9.89$--$11.99$\,km for the samples we adopt.
Substituting it for our J0614 channel shifts the inferred $R_{1.4}$ by $+0.06$\,km, well inside the statistical width of the posterior.
The assumed hotspot geometries of J0030 and J0614 are the largest single systematic among the four sources, so we investigate a sensitivity variant in \cref{sec:results:robustness} which swaps both to their alternative hotspot models, for J0030 the \texttt{ST+PDT} model of the earlier analysis \citep{Vinciguerra:2023qxq} and for J0614 the \texttt{ST+PDT} headline model of \citet{Mauviard:2025dmd}.
For J0614 that swap also changes the waveform resolution and the sampling settings, because the \texttt{ST+PDT} samples come from the converged headline run of \citet{Mauviard:2025dmd} whereas \texttt{PDT-U} was explored at lower resolution.
We do not include the fifth published NICER source, PSR~J1231--1411, whose inference converges only under a restricted radius prior and whose radius is mutually inconsistent with an independent reanalysis \citep{Salmi:2024bss,Qi:2025mpn}.

The mass of PSR~J0348+0432, long the benchmark massive-pulsar constraint at $2.01 \pm 0.04\,\Msun$ \citep{Antoniadis:2013pzd}, has been revised to $1.806 \pm 0.037\,\Msun$ by a CHIME/Pulsar remeasurement of the relativistic orbital decay that no longer relies on a white-dwarf atmosphere model \citep{Saffer:2024tlb}.
That value lies far below the support of our $\Mmax$ posterior, so the channel is inert and including or omitting it changes no digit we quote.
The massive-pulsar constraint enters instead through the NICER J0740 posterior, whose X-ray timing model carries the radio mass prior $2.08 \pm 0.07\,\Msun$ \citep{Fonseca:2021wxt}.

\section{Results}
\label{sec:results}

\subsection{Posterior EOS catalog}
\label{sec:results:catalogue}

Of the $1.2\times10^{9}$ parameter draws, two thirds lie in regions of the parameter box where pilot runs had shown the search never to succeed and are skipped (a random $1\%$ of them is still searched as a check, and a single one produced a valid \ac{EOS}, which has no measurable effect on the posterior).
The exponent existence search (\cref{sec:search}) rejects $98.6\%$ of the remaining $4.0\times10^{8}$ draws, i.e.\ for these draws no exponent vector rendering the \ac{EOS} causal and physical was found.
A further $13.5\%$ of the candidates is removed by the validity requirements of the subsequent \ac{TOV} stage, with practically no consequence for the posterior after the likelihoods have been applied.
They are either unphysical outright or lie at the compact extreme of the prior (\cref{sec:results:robustness}).
The crust-stability criterion of \cref{sec:search} then removes $10.3\%$ of the remaining candidates, and the symmetry-energy criterion a further $17.2\%$.
The resulting catalog holds $3\,631\,299$ \acp{EOS}, each with a combined likelihood weight $w_i$.
Under the full posterior weight the catalog retains a Kish \acl{ESS}, ${\rm ESS} = (\sum_i w_i)^2/\sum_i w_i^2$ \citep{Kish:1965}, of $2.9\times10^4$.
The two criteria of \cref{sec:search} that anticipate the finite-temperature tables hardly affect the inference.
The crust-stability criterion removes $10\%$ of the candidates but only about $4\%$ of the posterior mass, and the symmetry-energy criterion removes $17\%$ of the candidates and $16\%$ of the posterior mass.
Lifting either one moves the posterior medians quoted below by at most $0.02$\,km in $R_{1.4}$, $6$ in $\Lambda_{1.4}$, $0.002\,\Msun$ in $\Mmax$, $0.5$\,MeV in $L$ and $3$\,MeV in $K_{\rm sym}$.
Throughout this section, four weightings of the catalog appear under fixed names: prior (uniform weights over the catalog), only-nuclear (\ac{xeft} + \ac{pQCD}, \cref{sec:nuclear_likelihoods}), only-astro (NICER + GW170817, \cref{sec:astro}), and posterior (all likelihoods combined).

\begin{figure*}
  \centering
  \includegraphics[width=\textwidth]{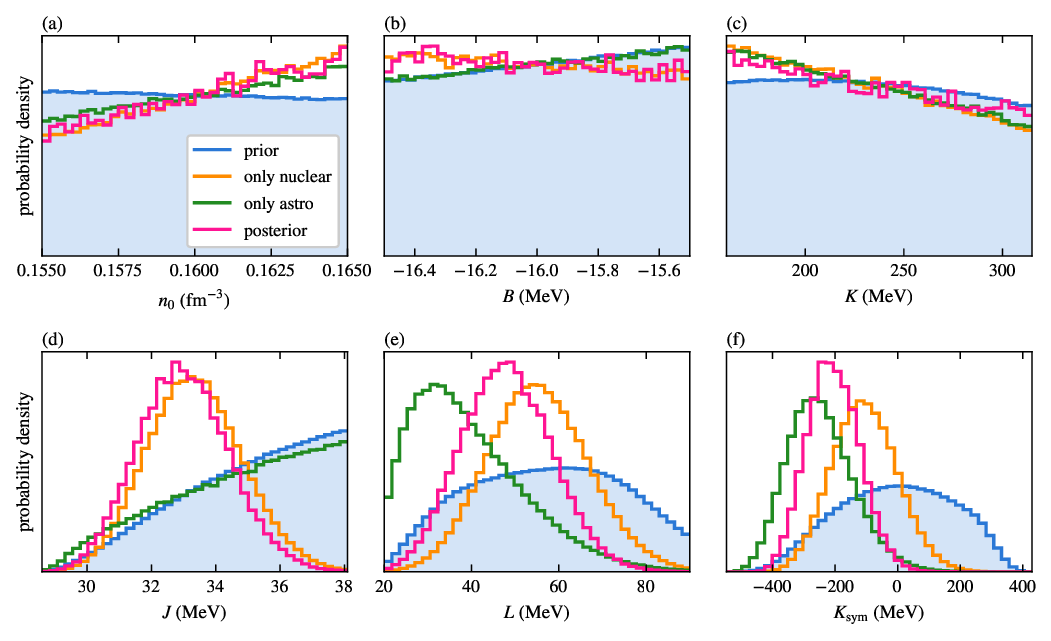}
  \caption{\label{fig:hist_saturation}
    Marginal distributions of the six saturation parameters under the prior (filled), only-nuclear, only-astro, and posterior weightings.
    Panels (a)--(c) ($n_0$, $B$, $K$): the reweighted marginals track the prior, whose gentle non-uniformity is the imprint of the exponent existence search and the two prior-level criteria of \cref{sec:search}, not of the box, which is uniform.
    $J$ (d) is narrowed by the chiral energy term.
    Panel (f) shows the literature coefficient $K_{\rm sym}$, not the sampled $\tilde{K}_{\rm sym}$ (\cref{sec:prior:saturation}).}
\end{figure*}
\Cref{fig:hist_saturation} shows the marginals of the six saturation parameters.
The \ac{SNM} parameters ($n_0$, $B$, $K$) largely reproduce their priors.
Both groups of likelihoods tilt the posterior slightly toward higher $n_0$ and lower $K$, but as anticipated in \cref{sec:prior:saturation}, they carry very little information on the \ac{SNM} sector.
The prior on $J$ rises toward the upper edge of its box, a consequence of the correlation with $L$ introduced through the unitary-gas constraint.
While the astrophysical data leave this shape nearly unchanged, the chiral energy term narrows $J$ to $33.0^{+2.4}_{-2.2}$\,MeV, well inside the box.
The other two neutron-rich-sector parameters, $L$ and $K_{\rm sym}$, are visibly reshaped as well: the slope $L$ tightens to $48^{+18}_{-16}$\,MeV (medians with $90\%$ credible intervals), and the symmetry-energy curvature $K_{\rm sym}$ is pulled toward softer values ($-208^{+143}_{-129}$\,MeV).
Note that panel (f) plots $K_{\rm sym}$ rather than the sampled $\tilde{K}_{\rm sym}$, because the two carry the same information and $K_{\rm sym}$ is the coefficient in which literature ranges are quoted.

While $J$ is almost exclusively constrained by the \ac{xeft} constraint, both groups of constraints contribute to the reshaping of $L$ and $K_{\rm sym}$.
The astrophysical data favor soft symmetry energies, while the nuclear-theory constraints exclude the softest values.
\begin{figure}
  \centering
  \includegraphics[width=\columnwidth]{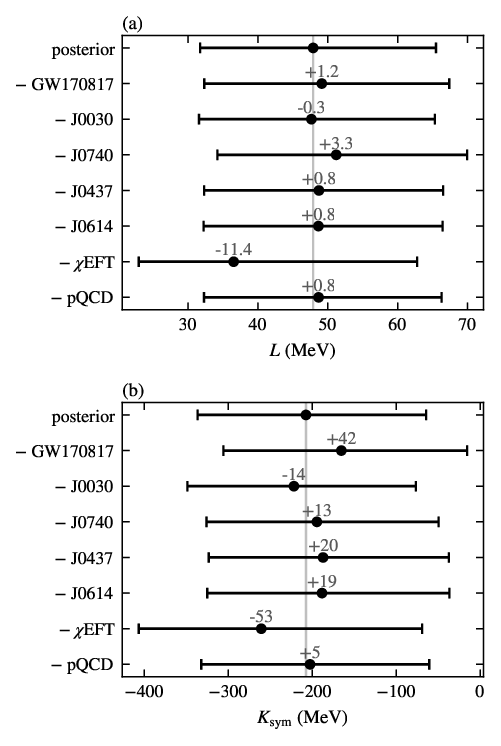}
  \caption{\label{fig:forest_knockout_sat}
    Knockout test for the symmetry-energy parameters $L$ and $K_{\rm sym}$: posterior medians and $90\%$ credible intervals after removing one likelihood at a time.
    The vertical line marks the full-posterior median; the number above each marker gives the median shift relative to it.}
\end{figure}
To isolate which specific constraint pulls which way, we remove one likelihood at a time from the posterior and record the resulting shift, shown for $L$ and $K_{\rm sym}$ in \cref{fig:forest_knockout_sat}.
Without the \ac{xeft} likelihood both posteriors move to markedly softer values, by $11$\,MeV in $L$ and $53$\,MeV in $K_{\rm sym}$, and the lower end of the $L$ interval drops from $32$ to $23$\,MeV.
The upper ends of both intervals barely change.
The \ac{xeft} likelihood thus closes the soft side of the symmetry energy, while the stiff side is limited by the astrophysical data.
Of its two parts, the pressure band carries this constraint, while the energy term alone moves $L$ and $K_{\rm sym}$ by less than $3$\,MeV and acts only on $J$.
Among the astrophysical data, every measurement except J0030 pulls toward softer values.
The largest shift in $L$ comes from J0740, and the largest shift in $K_{\rm sym}$ from GW170817, followed by the intermediate-mass pulsars J0437 and J0614.

\begin{figure}
  \centering
  \includegraphics[width=\columnwidth]{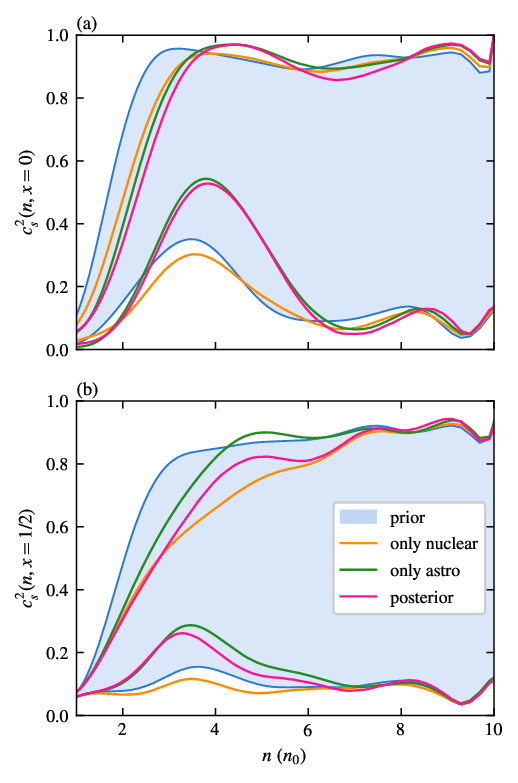}
  \caption{\label{fig:bands_cs_anchors}
    Squared speed of sound at fixed composition, including electrons, in (a) \ac{PNM} and (b) \ac{SNM}: $90\%$ credible bands for the prior (filled) and the only-nuclear, only-astro, and posterior weightings.
          }
\end{figure}
\Cref{fig:bands_cs_anchors} shows the squared speed of sound at fixed composition (\ac{PNM} and \ac{SNM}, the two channels in which the $c_s^2$ anchors of \cref{sec:prior:cs2} are sampled).
Although the $c_s^2$ anchors are sampled over the full causal interval $[0, 1]$ at every density, the prior is not uniform.
At saturation the sound speed is fixed by $K$ in \ac{SNM} and by $K + \tilde{K}_{\rm sym}$ in \ac{PNM}. In \ac{SNM} the electron contribution exceeds the nuclear one, so the full prior range of $K$ corresponds to a spread of less than $0.02$ in $c_s^2$.
Above $n_0$, a draw survives only where the \sro{} functional form can realize its $c_s^2$ targets given its saturation inputs with a valid \ac{EOS}, which means the speed of sound cannot rise arbitrarily fast.
However, in both panels the upper edge of the posterior band rises later than that of the prior.
The limit that the functional form places on how fast the sound speed can rise therefore lies outside the region the data select and does not shape the posterior.
This is what the exponent search of \cref{sec:search} was designed to achieve.
In tests, fixed or randomly drawn exponents produced a prior in which the speed of sound rises more slowly.
Draws with low speeds of sound produce very soft \acp{EOS} with a maximum-mass-\ac{NS} central density larger than $10\,n_0$, which are filtered out by the \ac{TOV} validity check after construction (they produce very small maximum \ac{TOV} masses and therefore do not contribute to the posterior).

\begin{figure}
  \centering
  \includegraphics[width=\columnwidth]{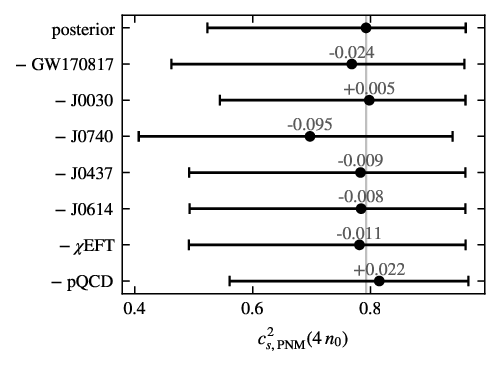}
  \caption{\label{fig:forest_knockout_cs}
    As in \cref{fig:forest_knockout_sat}, for the \ac{PNM} squared speed of sound at $4\,n_0$.}
\end{figure}

The likelihoods reshape the speed-of-sound bands in two ways.
Near saturation the sound speed of \ac{PNM} is tied to the symmetry-energy parameters $L$ and $\tilde{K}_{\rm sym}$.
The constraints that soften $L$ and $K_{\rm sym}$ in \cref{fig:hist_saturation} therefore also lower the \ac{PNM} sound speed there and remove the steepest rises below about $3\,n_0$.
The \ac{xeft} band does so directly, because $\beta$-equilibrated matter up to $1.5\,n_0$ is very neutron rich, and the astrophysical data do so because small radii and tidal deformabilities require a low pressure at $1$--$2\,n_0$.
An \ac{EOS} that is soft up to about $2\,n_0$ must then stiffen quickly to support a $2\,\Msun$ star.
The only-astro and posterior weightings accordingly lift the lower edge of the band around $3$--$4.5\,n_0$, to about $0.5$ in \ac{PNM}.
\Cref{fig:forest_knockout_cs} shows the posterior median and $90\%$ credible interval of the \ac{PNM} sound speed at $4\,n_0$ when one likelihood at a time is removed, and identifies the origin of this stiffening.
Removing J0740, the most massive of the four pulsars, lowers the median by $0.10$ and the lower end of the interval from about $0.5$ to $0.4$, far more than any other knockout.
Beyond $\sim6$--$7\,n_0$ the posterior relaxes back toward the prior.
Those densities lie above the central densities of most stars in the posterior, so the data provide no leverage there.
In principle \ac{pQCD} could provide leverage at these densities.
Its constraint requires that the \ac{EOS} can be connected to the perturbative regime at tens of saturation densities by a causal and thermodynamically consistent extension.
This condition tightens the higher the density at which it is imposed, so it would be strongest at our highest anchor, $10\,n_0$.
We nevertheless evaluate the \ac{pQCD} weight only at each \ac{EOS}'s own $\nmax$ (\cref{sec:nuclear_likelihoods}).
Imposing it at all densities above $\nmax$ would pull the posterior to lower speeds of sound there, but these densities are not realized in any stable star, so we do not weight the \ac{EOS} on them.

\begin{figure} [th!]
  \centering
  \includegraphics[width=\columnwidth]{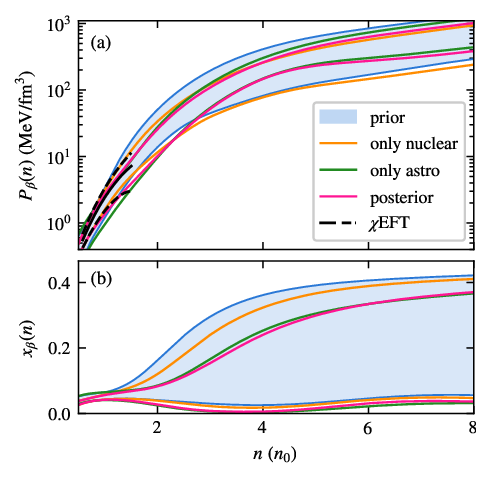}
  \caption{\label{fig:bands_P_xbeta}
    (a) $\beta$-equilibrium pressure and (b) proton fraction versus baryon density: $90\%$ credible bands for the prior (filled) and the only-nuclear, only-astro, and posterior weightings.
    Black lines: \ac{xeft} mean and $90\%$ credible band \citep{Gottling:2025ohe}, drawn over the reliable range of the N$^3$LO calculation ($n \le 1.5\,n_0$), which is also where the continuous \ac{xeft} likelihood of \cref{sec:nuclear_likelihoods} acts.}
\end{figure}
\Cref{fig:bands_P_xbeta} shows the $\beta$-equilibrium pressure $P(n, x_\beta)$ and proton fraction $x_\beta(n)$.
At low density, the pressure posterior is driven by the \ac{xeft} likelihood while the astrophysical constraints dominate the intermediate range of $2$--$3\,n_0$.
The high-density tail is softened by the \ac{pQCD} constraints, which we impose at each \ac{EOS}'s own $\nmax$ (\cref{sec:nuclear_likelihoods}).
However, the lower edge of the pressure band is set by the astrophysical constraints.
The $\beta$-equilibrium proton-fraction posterior is largely unconstrained by our data, as expected since the \ac{pQCD} and astrophysical constraints carry no direct information about the composition.
Only the \ac{xeft} likelihood tightens the band close to the saturation density.
At higher densities, the highest proton fractions in the prior are disfavored by the astrophysical constraints: the isovector sector is generally softer, so high proton fractions cannot support the $2\,\Msun$ pulsar masses.

\begin{figure} [th!]
  \centering
  \includegraphics[width=\columnwidth]{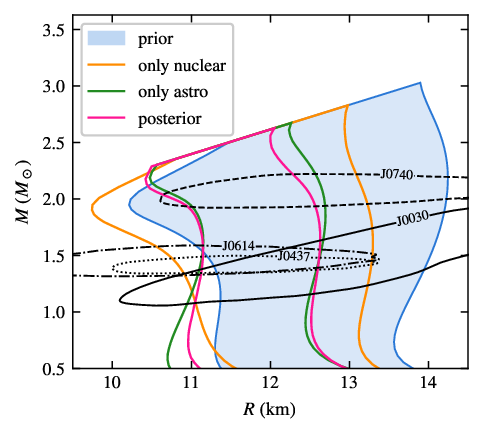}
  \caption{\label{fig:bands_MR}
    Mass--radius diagram: $90\%$ credible bands for the prior (filled) and the only-nuclear, only-astro, and posterior weightings.
    Thin black contours: $90\%$ highest-posterior-density regions of the four NICER $(M, R)$ posteriors entering the astrophysical likelihood.}
\end{figure}

\begin{figure*}
  \centering
  \includegraphics[width=\textwidth]{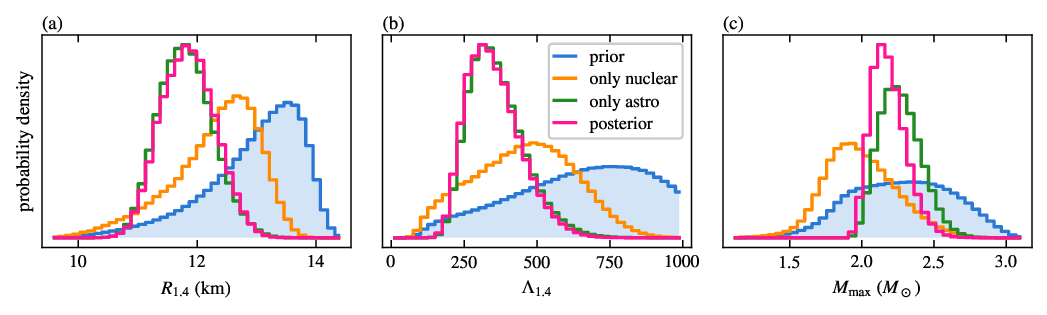}
  \caption{\label{fig:hist_observables}
    Marginal distributions of $R_{1.4}$, $\Lambda_{1.4}$, and $\Mmax$ under the prior, only-nuclear, only-astro, and posterior weightings.
    Axes are truncated; the prior's upper tail in $\Lambda_{1.4}$ extends beyond the plotted range.}
\end{figure*}

\begin{figure*}
  \centering
  \includegraphics[width=\textwidth]{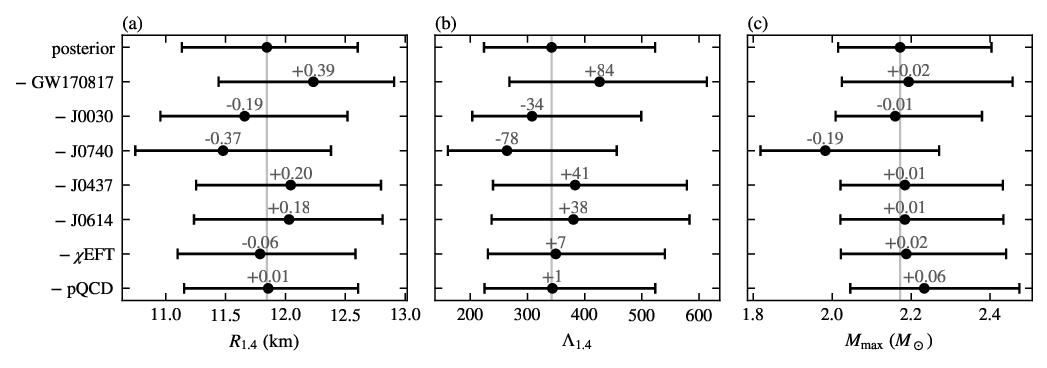}
  \caption{\label{fig:forest_knockout_obs}
    As in \cref{fig:forest_knockout_sat}, for the headline observables $R_{1.4}$, $\Lambda_{1.4}$, and $\Mmax$.}
\end{figure*}
We now turn to the stellar properties that follow from these \acp{EOS}.
\Cref{fig:bands_MR} shows the mass--radius bands, and \cref{fig:hist_observables} the marginal distributions of the three stellar observables $R_{1.4}$, $\Lambda_{1.4}$, and $\Mmax$.
The posterior yields $R_{1.4} = 11.84^{+0.76}_{-0.71}$\,km, $\Lambda_{1.4} = 342^{+181}_{-118}$ and $\Mmax = 2.17^{+0.23}_{-0.16}\,\Msun$ (medians with $90\%$ credible intervals).
\Cref{fig:forest_knockout_obs} shows the same three quantities for the knockout cases.
The radius and tidal polarizability at the $1.4\,\Msun$ reference mass are tightly correlated, so they largely follow the same trends.
Relative to the prior, the joint update shifts the $R_{1.4}$ and $\Lambda_{1.4}$ scales down and shrinks their $90\%$ width.
These shifts can mostly be attributed to GW170817 and the NICER measurements of J0437 and J0614.
The measurements of J0030 and J0740 mainly act on the lower end of radii, so they pull the overall posterior toward higher $R_{1.4}$ and $\Lambda_{1.4}$.
Meanwhile, the \ac{xeft} likelihood prefers larger radii, shifting the posterior to the right by a slight amount.
The maximum-mass posterior is most sensitive to the high-mass NICER pulsar J0740, whose removal lowers it, while \ac{pQCD} pushes it the other way and \ac{xeft} barely moves it.

Our median $R_{1.4}$ lies $0.2$ to $0.4$\,km below recent inferences combining \ac{xeft}, NICER and GW170817, which quote $R_{1.4} = 12.26^{+0.80}_{-0.91}$\,km \citep{Koehn:2024set} and $12.01^{+0.56}_{-0.75}$\,km for a speed-of-sound extension \citep{Rutherford:2024srk}, both at $95\%$ credibility.
These analyses predate the J0614 measurement, which lowers $R_{1.4}$ by $0.18$\,km in our analysis (\cref{fig:forest_knockout_obs}) and by about $0.1$\,km in \citet{Biswas:2025ivu}.
The remaining difference lies within the range spanned by the analysis variants of \cref{sec:results:robustness}.
\Citet{Mendes:2026mgc}, who include J0614 and use the same \ac{xeft} band, obtain a mass--radius posterior that covers the same radius range as ours at $1.4\,\Msun$, and find $L = 44.2$--$56.7$\,MeV at $68\%$ credibility for a speed-of-sound extension, consistent with our $48^{+18}_{-16}$\,MeV at $90\%$.

\subsection{Robustness}
\label{sec:results:robustness}

\begin{figure*}
  \centering
  \includegraphics[width=\textwidth]{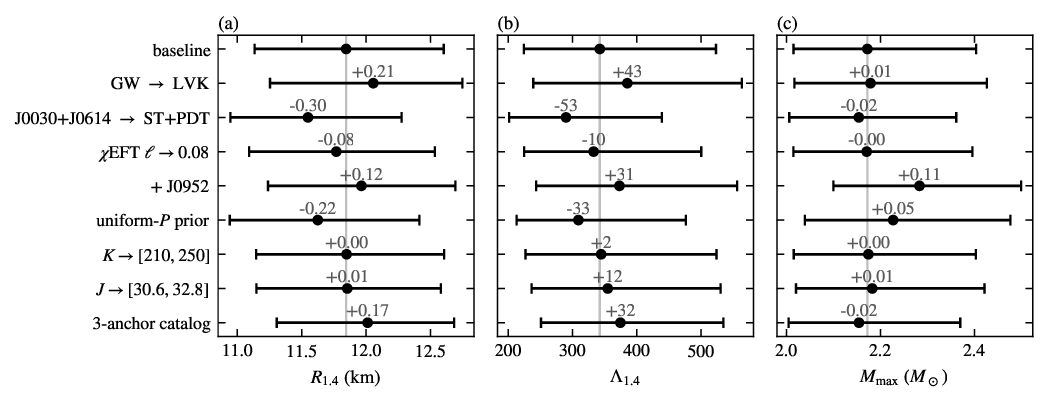}
  \caption{\label{fig:forest_sensitivity}
    Sensitivity test: posterior medians and $90\%$ credible intervals under one-at-a-time swaps of the analysis inputs, relative to the baseline (top row; vertical line).
    From top to bottom, the swaps are the LIGO--Virgo GW170817 posterior \citep{LIGOScientific:2018hze} in place of ours, the \texttt{ST+PDT} hotspot models of J0030 and J0614 (for J0614 this also changes the waveform resolution and sampling settings, see \cref{sec:astro}), the doubled \ac{xeft} correlation length $l = 0.08\,{\rm fm^{-3}}$, the addition of the black-widow pulsar J0952, a prior uniform in the $\beta$-equilibrium pressures at $\{2,4,6\}\,n_0$, a restriction of the $K$ box to the giant-monopole consensus $[210,250]$\,MeV, a restriction of the $J$ box to the \ac{xeft} range $[30.6,32.8]$\,MeV, and the companion three-anchor catalog.
    The number above each marker gives the median shift relative to the baseline.
  }
\end{figure*}

In this subsection we investigate the dependence of the posterior shape on the analysis choices we lay out in \cref{sec:model,sec:likelihoods}.
To this end, we swap one ingredient of the analysis pipeline at a time and report the resulting shifts in the headline observables $R_{1.4}$, $\Lambda_{1.4}$, and $\Mmax$ in \cref{fig:forest_sensitivity}.
The variations fall into three groups.
For the astrophysical likelihoods, we replace our GW170817 posterior with the low-spin-prior posterior of the LIGO--Virgo Collaboration, swap the NICER J0030 and J0614 hotspot models from \texttt{PDT-U} to \texttt{ST+PDT}, and add the black-widow pulsar J0952.
We keep J0952 as a variant rather than a baseline channel because of how its mass is measured.
The baseline massive-pulsar constraint rests on the Shapiro delay of J0740, a geometric effect in the radio timing that is largely independent of any model of the companion \citep{Fonseca:2021wxt}.
The mass of J0952 instead follows from modeling the optical light curve and radial velocities of its irradiated companion, which requires a heating model and the orbital inclination \citep{Romani:2022jhd}.
\Citet{Romani:2022jhd} argue that the modeling uncertainty is small for this system, since the companion stays well inside its Roche lobe and a simple direct-heating model fits best, and we have no reason to dispute that.
We nonetheless treat it as a variant because it is a single system measured by a single method, its uncertainty of $\pm0.17\,\Msun$ is larger than the $\pm0.07\,\Msun$ of J0740, and its central value lies above every Shapiro-delay mass, so promoting it to the baseline would make one system the dominant constraint on $\Mmax$.
The nuclear-physics inputs are probed by doubling the \ac{xeft} correlation length to $l = 0.08\,{\rm fm^{-3}}$.
The prior construction is varied in three ways.
We reweight the prior from uniform in the sampled $c_s^2$ targets to uniform in the $\beta$-equilibrium pressure, we switch the speed-of-sound anchor scheme from five to three anchors, and we restrict the $K$ and $J$ saturation-parameter boxes to their literature-consensus intervals.
Each swap except the anchor scheme is applied in isolation on the same catalog, so the reported shift isolates a single analysis choice.

The LIGO--Virgo posterior favors larger reduced tidal polarizabilities than ours, and adopting it raises the median $R_{1.4}$ by $0.21$\,km and $\Lambda_{1.4}$ by $43$, while $\Mmax$ is unchanged.
This shift is comparable to those from the NICER hotspot models and the prior choice and well inside the $90\%$ credible interval, so the waveform model and the parameter-estimation setup of the GW analysis are not a dominant systematic.
The anchor-scheme test of \cref{sec:prior:cs2} evaluates the full posterior on the companion three-anchor catalog of $2.99\times10^6$ \acp{EOS}.
That catalog retains a Kish \ac{ESS} of $5.8\times10^4$ under the full posterior weight, larger than that of the five-anchor catalog because the coarser grid concentrates the weight less.
The comparison is therefore not limited by the companion's statistics.
It yields slightly higher $R_{1.4}$ and $\Lambda_{1.4}$, by $0.17$\,km and $32$.
This shift reflects the reduced flexibility of the coarser grid: with five anchors, the pressure can stay soft around $2\,n_0$, where $R_{1.4}$ is set, and rise steeply only at higher densities, so an \ac{EOS} can reach low $R_{1.4}$ while still fulfilling the $2\,\Msun$ pulsar constraints.
The three-anchor form ties these density regions together and excludes such late-stiffening solutions.
This shows that the large number of degrees of freedom in the supranuclear regime is needed to keep modeling assumptions from limiting the posterior.
An intermediate four-anchor scheme, which we do not report in detail here, moved the median and the $90\%$ interval of $R_{1.4}$ by at most $0.1$\,km relative to five anchors.
The joint swap of the J0030 and J0614 hotspot models shifts the $R_{1.4}$ posterior by $-0.30$\,km.
All remaining variants shift the posterior by comparatively small amounts: a prior reweighted to be uniform in the $\beta$-equilibrium pressures at $\{2,4,6\}\,n_0$ instead of uniform in the sampled $c_s^2$ targets ($-0.22$\,km), the \ac{xeft} correlation length doubled to $l = 0.08\,{\rm fm^{-3}}$ ($-0.08$\,km), and adding the fast-spinning black-widow pulsar J0952--0607 \citep{Romani:2022jhd}, which mainly hardens the $\Mmax$ constraint ($+0.11\,\Msun$).
The $\Mmax$ posterior is most sensitive to the addition of the black-widow pulsar J0952, which is measured to be heavier than J0740 and therefore shifts the distribution up.
Restricting the catalog to the giant-monopole consensus $K \in [210, 250]$\,MeV shifts the $R_{1.4}$ posterior median by less than $0.01$\,km, confirming that the wide $K$ box of \cref{sec:prior:saturation} carries no constraining power of its own.
The analogous restriction of the $J$ box to the \ac{xeft}-based region $J = 31.7 \pm 1.1$\,MeV from \citet{Drischler:2020yad} shifts the $R_{1.4}$ median by only $0.01$\,km, likewise small against the width of the posterior.
Taken together, the variants move the medians by $-0.30$ to $+0.21$\,km in $R_{1.4}$, $-53$ to $+43$ in $\Lambda_{1.4}$ and $-0.02$ to $+0.05\,\Msun$ in $\Mmax$, apart from the black-widow pulsar J0952, which raises $\Mmax$ by $0.11\,\Msun$.

Restricting our \ac{EOS} catalog to be valid up to $10\,n_0$ also carries ample margin over the densities actually realized: the $\nmax$ distribution of the catalog \acp{EOS} with $\Mmax > 1.97\,\Msun$ has median $5.0\,n_0$, $90$th percentile $6.6\,n_0$, and $99.9$th percentile $8.3\,n_0$, and the posterior stays within the same range ($99$th percentile $7.5\,n_0$).

A final robustness check concerns the astrophysical importance estimator itself (\cref{app:estimator}), which carries Monte-Carlo noise from the finite \ac{KDE} training, query-mass, and mass-marginal subsamples and depends on the kernel bandwidth.
We set the bandwidth by Scott's rule, the standard $N^{-1/(d+4)}$ scaling with sample size $N$ in $d$ dimensions \citep{Scott:1992}.
We quantify both the Monte-Carlo noise and the bandwidth dependence.
Recomputing the five astrophysical weights of the headline posterior under three independent seeds on a tenth of the catalog, the per-\ac{EOS} log-weight scatter has a median of $0.03$--$0.12$, depending on the channel.
This scatter is measured after removing the overall normalization of each seed, which is common to all \acp{EOS} and cancels in every posterior quantity.
These fluctuations largely average out in the reweighted quantiles, leaving a seed-to-seed spread of the posterior $R_{1.4}$ and $\Mmax$ medians of $0.02$\,km and $0.003\,\Msun$ respectively.
Varying the kernel bandwidth by $\pm20\%$ about the Scott value (at fixed seed) shifts those medians by at most $0.02$\,km and $0.001\,\Msun$.
Both effects are more than an order of magnitude below the width of the posterior.

\section{Selection of a simulation ensemble}
\label{sec:selection}

For use in simulations we select a small ranked ensemble of \acp{EOS}.
The ensemble should bracket the \ac{EOS}-induced posterior uncertainty of merger observables such as ejecta masses and composition, waveforms and postmerger spectral peaks, so that the spread across simulations run with its members reflects that uncertainty.
At the same time, every member should remain individually plausible under the posterior of the previous section.
We assume that the $\beta$-equilibrium pressure is the \ac{EOS} property with the largest influence on these observables, since it determines the structure of the stars during the inspiral and, to a good approximation, that of the merger remnant.
We therefore characterize each \ac{EOS} first by $\log P_\beta(n)$ at the seven density anchors $n/n_0 \in \{1, 1.5, 2, 3, 4, 5.5, 6.5\}$.

Beyond the pressure, the dependence of the \ac{EOS} on the proton fraction enters through two further quantities.
The first is motivated by the postmerger neutrino emission of a long-lived remnant.
The merger heats matter that is still as neutron rich as in the cold stars, and the equilibrium proton fraction of hot matter is higher.
Such a remnant therefore protonizes through an excess of electron antineutrinos over electron neutrinos, sourced by positron captures at densities around and above saturation \citep{Sekiguchi:2015dma, Perego:2019adq}.
The luminosity and mean-energy ratios of the two species set the electron fraction of the neutrino-driven and spiral-wave winds \citep{Qian:1996xt}, which dominate the ejecta of long-lived remnants on the timescale of hundreds of milliseconds and whose nucleosynthesis yields and kilonova signal depend sensitively on the electron fraction in exactly this range \citep{Bernuzzi:2024mfx, Jacobi:2025eak}.
How far the remnant has to protonize depends on the initial proton fraction of its matter, the cold equilibrium value $x_\beta(n)$.
The neutrinos themselves decouple far below saturation, and those produced above it reach the surface by diffusion over hundreds of milliseconds.
The \ac{xeft} band determines $x_\beta$ around and below $n_0$, so the remaining freedom lies above.
We therefore add the cold equilibrium proton fraction at $n = 2\,n_0$, which sets the rise of the proton fraction across the outer core of the remnant, the region from which neutrinos diffuse out earliest.
We expect this effect to be smaller than that of the pressure, and the proton fraction accordingly enters the selection with a smaller weight (see below).
The second quantity beyond the pressure is the neutron-matter energy $E_{\rm PNM}(0.10\,{\rm fm^{-3}})$, which the chiral energy term of \cref{sec:nuclear_likelihoods} constrains directly.
The pressure anchors fix the slope of the energy per particle but not its value, and without this axis the selection scatters the members across the full prior range of the sub-saturation neutron-matter energy, far outside the chiral band, while their pressures agree.

We reduce the standardized $\log P_\beta(n)$ anchors by a posterior-weighted principal-component analysis.
Four principal components are enough to capture $98.1\%$ of the weighted variance.
The four principal $P_\beta$ components are complemented by the z-scored $x_\beta(2\,n_0)$ and $E_{\rm PNM}(0.10\,{\rm fm^{-3}})$, resulting in a six-dimensional feature space.

We build our final \ac{EOS} ensemble in two steps.
First, we build a weighted \ac{KDE} of the catalog in this feature space and take the catalog \ac{EOS} nearest to the posterior-weighted mean of the features as our fiducial choice.
Second, we cut the catalog to only the plausible members, which we define as the highest-posterior-density region \citep[e.g.][]{Gelman:2013bda} of the \ac{KDE} distribution containing $32\%$ of the total posterior mass (the selection region)\footnote{This might sound like an extremely narrow cut but only because most of the mass of a six-dimensional distribution sits far from its center. For the one-dimensional marginals of the distribution this cut is roughly equivalent to a standard $95\%$ credible interval.}.
Then a farthest-point (max--min) greedy traversal of that region \citep{Gonzalez:1985} produces the ranked list.
Every catalog \ac{EOS} inside the region is a candidate, and no further weight threshold is applied.
The first member is the fiducial choice.
Each subsequent member is the catalog point that maximizes the Euclidean distance in the feature space to its nearest already-selected neighbor.
Since we retain the principal components at their eigenvalue scaling rather than sphered to unit variance, the distance in the feature space is variance-weighted and the most varying principal component is traversed preferentially.
The two z-scored dimensions have unit variance by definition, below the three leading $P_\beta$ components (eigenvalues $2.5$, $2.1$, and $1.7$) though above the fourth ($0.5$).
Altogether the pressure carries about $80\%$ of the variance in the feature space, and the other two quantities about $10\%$ each.
This is intended, since we expect the influence of the composition and of the low-density neutron-matter energy on the simulation observables to be subdominant with respect to the main stiffness modes.
The ranking is nested, so a simulation campaign can use a truncated selection or be extended by more \acp{EOS} later.

\begin{table*}
  \centering
  \caption{\label{tab:selection}
    Properties of the twelve ranked ensemble members (rank 1 is the fiducial): incompressibility $K$, symmetry-energy parameters $J$, $L$, and $K_{\rm sym} = 9 n_0^2\, \mathrm{d}^2 E_{\rm sym}/\mathrm{d}n^2|_{n_0}$, where $E_{\rm sym}(n) = \veps(n, 0) - \veps(n, 1/2)$; the $\beta$-equilibrium sound speed at $2$ and $5\,n_0$ and proton fraction at $2\,n_0$; and \ac{TOV} observables.
    The last two columns give the neutron-matter energy $E_{\rm PNM}(0.10\,{\rm fm^{-3}})$ that enters the chiral likelihood and the member's own posterior weight relative to the fiducial, $w/w_0$.
  }
  \begin{tabular}{rccccccccccccc}
    rank & $K$ & $J$ & $L$ & $K_{\rm sym}$ & $c_s(2n_0)$ & $c_s(5n_0)$ & $x_\beta(2n_0)$ & $R_{1.4}$ & $\Lambda_{1.4}$ & $\Mmax$ & $\Rmax$ & $E_{\rm PNM}$ & $w/w_0$ \\
     & [MeV] & [MeV] & [MeV] & [MeV] & & & & [km] & & [$\Msun$] & [km] & [MeV] & \\
    \hline
    \input{selection_table}
  \end{tabular}
\end{table*}

\begin{figure*}
  \centering
  \includegraphics[width=\textwidth]{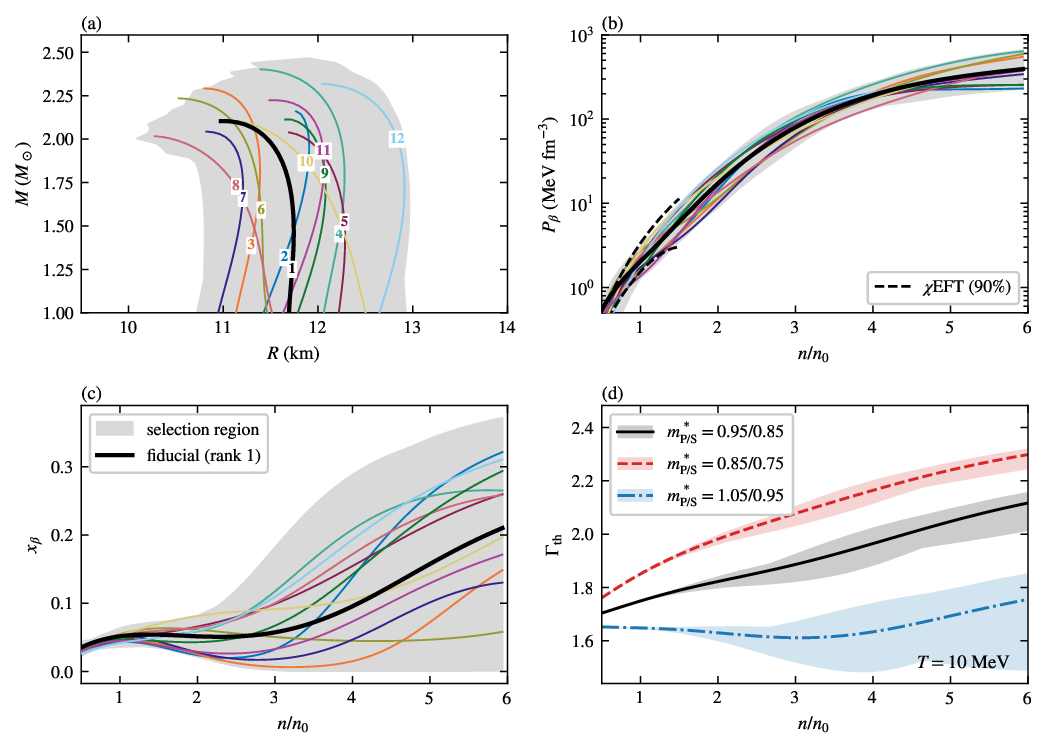}
          \caption{\label{fig:selection_mstar}
    The delivered simulation ensemble: the fiducial member (rank 1) and the eleven further ranked members, over the full support of the selection region of \cref{sec:selection} (gray: the envelope, at each mass or density, of every catalog member inside the region).
    Panels: (a) mass--radius curves, (b) cold $\beta$-equilibrium pressure $P_\beta(n)$, (c) equilibrium proton fraction $x_\beta(n)$, and (d) the thermal index $\Gamma_{\rm th}(n)$ at $T = 10$\,MeV.
    The dashed black lines in panel (b) mark the $90\%$ credible band of the \ac{xeft} calculation \citep{Gottling:2025ohe}, as in \cref{fig:bands_P_xbeta}.
    In panels (a)--(c) the fiducial (rank 1) is the thick black curve and the remaining members are rank-colored; each curve is numbered by its rank, inline in panel (a) only.
    Panel (d) shows the three delivered effective-mass levels as linestyle ($\pm0.10$ about the sector-split reference $m^*_{\rm PNM}/m^*_{\rm SNM} = 0.95/0.85$). Each line is the fiducial member at that level and each shaded band is the envelope over all twelve members, which the split (composition-dependent) effective masses render non-negligible ($\Delta\Gamma_{\rm th} \sim 0.09$--$0.28$ across the ensemble).}
\end{figure*}
We deliver twelve members, the fiducial and eleven ranked alternatives shown in \cref{fig:selection_mstar}.
Panel (a) shows their $M$--$R$ diagrams, panel (b) their $P_\beta(n)$ curves.
Their own posterior weights, $w/w_0$ in \cref{tab:selection}, lie between a fifth and eight times the fiducial's for ranks 1 to 11, while the twelfth member, the stiffest, sits at $0.9\%$ of the fiducial weight, at the edge of the selection region.
The first pick, the fiducial \ac{EOS}, sits at $R_{1.4} = 11.74$\,km, $\Lambda_{1.4} = 317$, $\Mmax = 2.10\,\Msun$.
Gray bands show the full pointwise support of the selection region, so every pick lies inside it by construction.
\Cref{tab:selection} lists each member's saturation parameters, intermediate-density properties, and \ac{TOV} observables.

\begin{figure}
  \centering
  \includegraphics[width=\columnwidth]{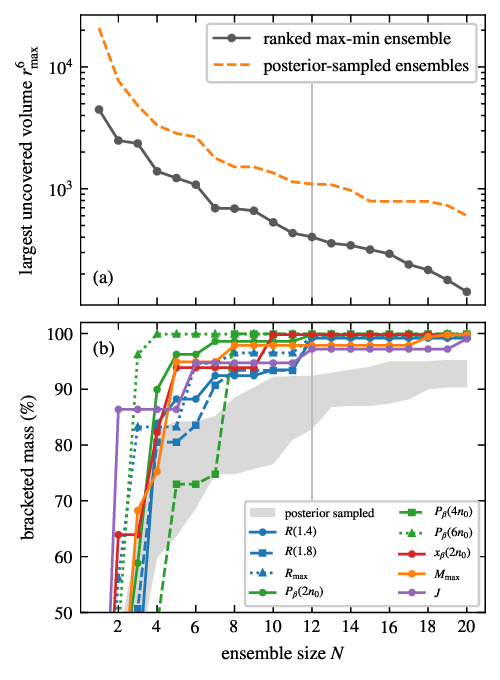}
  \caption{\label{fig:selection_bracketing}
    Convergence of the ranked ensemble with the number of delivered members $N$.
    (a) The volume $r_\mathrm{max}^6$ of the largest member-free ball in the variance-scaled six-dimensional selection space, for the ranked max--min ensemble and for ensembles of $N$ random posterior draws (median over 16 realizations).
    (b) The share of the posterior mass in the selection region whose observable value lies between the member minimum and maximum, for each labeled observable.
    Catalog \acp{EOS} for which an observable is undefined, e.g.\ $P_\beta(6\,n_0)$ when the table ends below $6\,n_0$, count as not bracketed.
    The gray band spans, over the same observables, the bracketed mass of the random posterior-drawn ensembles (median over 16 realizations).
    The vertical line marks the delivered $N = 12$.}
\end{figure}
\Cref{fig:selection_bracketing} quantifies the convergence of this construction with the truncation depth $N$ and motivates the delivered ensemble size.
Panel (a) follows the max--min objective itself.
The coverage radius $r_{\rm max}$ is the largest distance in the feature space from any plausible catalog \ac{EOS} to its nearest ensemble member, and we plot the corresponding volume $r_{\rm max}^6$, which for an evenly filled region falls as $1/N$ independent of the dimension.
The dashed curve shows the same statistic for ensembles of $N$ \acp{EOS} drawn at random from the posterior, as the median over 16 realizations.
For every $N \ge 2$ the ranked ensemble leaves a volume $2.0$ to $4.2$ times smaller, and the random curve reaches the coverage of the ranked ensemble at $N = 12$ only around $N \approx 22$, beyond the plotted range.
The measured volume falls by a factor of $17$ between $N = 2$ and $20$, somewhat faster than the factor of $10$ of an even filling.
It decreases steadily and thus provides no obvious point of truncation.

Panel (b) instead scores the same nested ensembles observable by observable, in terms of the physical quantities the ensemble is meant to bracket.
We score each observable separately because a dozen members cannot enclose a six-dimensional region, whereas they can bound every one-dimensional marginal from both sides.
For each observable and each depth $N$ we take the minimum and the maximum over the first $N$ members as the bracket and sum the posterior mass of all plausible catalog \acp{EOS} whose value falls inside it.
A bracketed mass of $100\%$ means simulations of the $N$ members bound the plausible range of that observable from both sides, while the missing mass corresponds to \acp{EOS} more extreme than every member.
In this view natural cut-offs do appear.
Four members already bracket $P_\beta(6\,n_0)$ to $99.9\%$ and $P_\beta(2\,n_0)$ to $90\%$, while the proton fraction, the radii, $\Mmax$ and $P_\beta(4\,n_0)$ lag behind at $48$--$84\%$.
The fifth member lifts $\Mmax$, $x_\beta(2\,n_0)$ and $P_\beta(2\,n_0)$ to $94$--$96\%$, the eighth, a soft member with $\Mmax = 2.02\,\Msun$ and $R_{1.4} = 11.37$\,km, lifts $P_\beta(4\,n_0)$ from $75\%$ to $99.9\%$ and $\Mmax$ to $97.9\%$, and the tenth closes the proton fraction at $99.8\%$.
The twelfth member, the stiffest with $R_{1.4} = 12.85$\,km and $\Lambda_{1.4} = 609$, lifts the radii and $P_\beta(2\,n_0)$ from $93$--$99\%$ to above $99\%$.
We therefore deliver $N = 12$ members, at which point every observable in \cref{fig:selection_bracketing} is bracketed at the $97\%$ level or better.
The symmetry energy $J$ is bracketed to $86\%$ by two members, to $95\%$ from the sixth and to $97\%$ from the twelfth, above the random draws at every $N$.
Beyond twelve the curves are flat, except for $\Mmax$, which reaches $99.4\%$ at $N = 18$, and $J$, which reaches $99\%$ at $N = 20$.
Across the twelve members the ensemble spans $R_{1.4} = 11.13$--$12.85$\,km, $\Lambda_{1.4} = 239$--$609$, $\Mmax = 2.02$--$2.40\,\Msun$, and $x_\beta(2\,n_0) = 0.024$--$0.081$, covering the bulk of the posterior $90\%$ interval of each observable.
Applications needing fewer members can truncate the nested ranking at eight, and later campaigns can extend it without invalidating earlier runs.
The member list itself is not unique.
The max--min picks sit on the boundary of the selection region, and that boundary is estimated from a random subsample of the catalog.
Repeating the selection with four different subsamples returns the same fiducial but eleven different further members every time.
The brackets they span agree to within $0.3$\,km in $R_{1.4}$, $0.06\,\Msun$ in $\Mmax$ and $0.007$ in $x_\beta(2\,n_0)$, and the bracketed mass of each observable at $N = 12$ changes by at most $9$ percentage points, for $\Mmax$, and by $1$ to $4$ points for the pressures, radii and proton fraction.
The delivered ensemble is therefore one representative of a class of ensembles that bracket the posterior comparably well.
It also holds up under the analysis variants of \cref{sec:results:robustness}.
Re-scoring the fixed twelve members against each variant posterior, every radius, pressure, proton-fraction and $\Mmax$ marginal stays bracketed at $81$--$96\%$ of that posterior's mass, against $90$--$95\%$ for the baseline when counted in the same way (as a share of the full posterior and not of the selection region).

The catalog fixes the effective masses at the reference values of \cref{sec:model} for every member, so the selection above ranks cold-sector diversity only and none of the data in this analysis directly constrain the nucleon effective mass, which controls the thermal response of the \ac{EOS}.
We therefore augment each selected member with effective-mass variants that change the thermal response at fixed cold-sector behavior.
Since the \sro{} effective mass cannot follow the density dependence predicted by \ac{xeft} \citep{Keller:2020qhx,Keller:2022crb}, neither the reference level nor the variants are consistent with the \ac{xeft} thermal response, and the variants do not represent an \ac{xeft} uncertainty band.
We provide them nonetheless because they enable a systematic study of how strongly simulation outcomes depend on the thermal sector.
The variants we supply are a low (strong-thermal) level at $m^*_{\rm SNM}/m = 0.75$, $m^*_{\rm PNM}/m = 0.85$ and a high (weak-thermal) level at $0.95$, $1.05$.
With the reference level, each member is thus released at three effective-mass levels, for a total of 36 tables.
The member's full parameter set, the 16 cold-sector inputs (six saturation parameters and ten speed-of-sound targets) together with the exponent vector $\delta$, is held fixed, and only the effective-mass targets are shifted.
This modification preserves the cold $\beta$-equilibrium pressure $P_\beta(n)$ and composition $x_\beta(n)$ to good accuracy.
Across all twelve members the variant tables shift $R_{1.4}$ by at most $17$\,m, $\Mmax$ by at most $0.006\,\Msun$, and the cold $\beta$-equilibrium pressure by at most $2\%$ between $n_0$ and $8\,n_0$.
The resulting spread in the thermal index $\Gamma_{\rm th}$ (\cref{eq:gammath}) at $k_B T = 10$\,MeV and $x=x_\beta$ is shown in panel (d) of \cref{fig:selection_mstar}.
The spread within each $m^*$ variant comes from the proton-fraction dependence of the thermal index through the member-to-member variation of $x_\beta$.

Each released \ac{EOS}, the twelve ensemble members and their effective-mass variants, is propagated with the \sro{} code into a complete general-purpose finite-temperature table, tabulated in baryon density, temperature, and electron fraction $Y_e$ (which is equivalent to $x$) on a grid of $400 \times 200 \times 96$ points.
The temperature spans $0.01$--$300$\,MeV and the proton fraction $0.01$--$0.6$ in every table, reaching proton-rich matter, with uniform spacing in $\log_{10} T$ and in $Y_e$.
For the member ranked third, the cold $\beta$-equilibrium composition falls below $Y_e = 0.01$ near $3\,n_0$, so its three tables carry one additional row at $Y_e = 0.0038$ and have $97$ proton-fraction points, which keeps the $\beta$-equilibrium track inside the grid for every member.
The density range is tailored per table and runs from $10^{-12}\,{\rm fm^{-3}}$ up to one grid spacing below that \ac{EOS}'s own construction validity limit, the lowest density at which monotonicity or causality fails ($10.2\,n_0$ for the fiducial).
The table check additionally requires the electron-inclusive pressure to rise monotonically on every fixed-$Y_e$ slice up to $Y_e = 0.51$ and along the $\beta$-equilibrium track, and it lands between $9.2$ and $10.2\,n_0$ for all twelve members.
No table is cut by the sound-speed check, and every table extends to at least $1.4\,\nmax$.

Each table is assembled from two \sro{} solves sharing the member's Skyrme coefficients.
A \acf{SNA} solve covers the full density range and describes inhomogeneous matter as a dense phase, representing nuclei, coexisting with a dilute gas of alpha particles and unbound nucleons, with the dissolution of nuclei into uniform matter (crust melting) treated as a first-order phase transition.
This solve leaves defects at the boundary of the inhomogeneous phase, which we repair as described in \cref{app:sna_repair}.
In the repaired tables the pressure decreases with density in at most $0.01\%$ of the cells, no cell has an energy that decreases with temperature, and the sound speed lies between zero and the speed of light everywhere.
A companion \acs{NSE} solve covers the low-density regime with an ensemble of 3332 nuclides whose masses and partition functions come from the JINA REACLIB compilation \citep{Cyburt:2010}.
The two solutions are blended in the free energy with a smooth $\tanh$ weight in $\log_{10} n$, centered at a transition density $n_t = 10^{-3}\,{\rm fm^{-3}}$ with a width of $0.33$\,dex, one decade above the \sro{} default so that the dilute region in which the single-nucleus solve is least reliable is handed to the \ac{NSE} ensemble.
Below the transition the \ac{NSE} ensemble with its experimental masses is the more faithful description, while above it the \ac{SNA} treatment, derived from the member's own functional, better captures the increasingly neutron-rich matter approaching neutron drip and the crust-core transition.

Along cold $\beta$-equilibrium the delivered tables reproduce the baryonic pressure of the underlying functional to better than $0.15\%$ everywhere above their crust-core transition and to $0.05\%$ between $1$ and $5\,n_0$.
Integrating the \ac{TOV} equations on the tables' own cold slices, with their \ac{SNA}/\ac{NSE} crust in place of the BPS crust used in the catalog, changes $R_{1.4}$ by at most $0.04$\,km ($0.3\%$), $R(M)$ by at most $0.5\%$ above $1\,\Msun$, and $\Mmax$ by at most $0.004\,\Msun$.
The radius difference is largest at $1\,\Msun$ and shrinks with mass, since the crust makes up a larger share of the radius in lighter stars.
Over the full three-dimensional grid, the pressure and entropy agree with the derivatives of the free energy per baryon $f$, $P = n^2 (\partial f/\partial n)_{T,Y_e}$ and $s = -(\partial f/\partial T)_{n,Y_e}$, and the Maxwell relation between them holds, to better than $0.3\%$ in $95\%$ of the cells, evaluated with centered finite differences on the table grid.
Leaving out the $1$ to $2\%$ of cells whose difference stencil straddles the edge of the nuclear region, fewer than $0.5\%$ of the cells deviate by more than $1\%$ in any table.

\section{Conclusions}
\label{sec:conclusions}

We have presented an extensive multimessenger inference of the nuclear \ac{EOS} employing modern constraints from nuclear physics and astrophysics, based on the simplified Skyrme functional of the \sro{} code.
Based on the resulting \ac{EOS} catalog, we introduce a general and systematic method for reducing a weighted multimessenger \ac{EOS} posterior to a small, ranked ensemble of finite-temperature \acp{EOS} ready for simulation.
The construction is deliberately generic: the selection of \cref{sec:selection} takes any weighted \ac{EOS} catalog as input and applies regardless of how that posterior was produced, so it is not tied to the Skyrme sampling used here.
With the large simulation campaigns that a new generation of GPU-accelerated, exascale-capable numerical-relativity codes is now making feasible, our catalog enables the direct and systematic propagation of nuclear- and astrophysics-derived constraints to the predictions of observables by numerical simulations of compact-object mergers or supernovae, such as the gravitational-wave signal, the nucleosynthesis, or the electromagnetic transients.

The works closest to the one presented here are the extended-Skyrme Bayesian inferences of Beznogov and Raduta \citep{Beznogov:2023jqp,Beznogov:2024vcv}, together with their families of finite-temperature tables \citep{Raduta:2025yst,Raduta:2025fpr}, and the finite-temperature free-energy model of \citet{Du:2021rhq}.
Both programs deliver small sets of simulation-ready general-purpose tables, and both differ from this work in where the data act on the \ac{EOS} and in how the released tables relate to the inference.
Neither includes the multimessenger constraints from GW170817 and NICER.
In the Beznogov--Raduta inference every constraint acts, by deliberate design, at or near saturation density (nuclear empirical parameters together with \ac{xeft} \ac{PNM} energies and effective masses at $n \le n_0$), with astrophysics entering only as a hard maximum-mass cut, and their own analysis concludes that the high-density behavior of the posterior is dominated by the functional rather than by the constraints \citep{Beznogov:2023jqp}.
\citet{Du:2021rhq} do constrain the high-density sector observationally, but through X-ray mass--radius analyses predating GW170817 and NICER, entering as a discrete set of samples that, as the authors note, disfavors strong phase transitions.
In the relativistic mean-field tables of \citet{Alford:2023rgp} the high-density behavior is likewise set by the functional.
Their couplings are fitted to \ac{xeft} neutron matter below $2\,n_0$ and to saturation properties \citep{Alford:2022bpp}, and the behavior at higher densities follows entirely from these fits, so the spread of the four released models there is the low-density uncertainty extrapolated through the functional form rather than an independent degree of freedom.
In our approach the high-density degrees of freedom are sampled separately from the saturation and \ac{xeft} sector and are constrained directly by the astrophysical and \ac{pQCD} data, which avoids such model-imposed correlations between the two regimes.
The exponent search in addition keeps every \ac{EOS} causal at fixed composition up to $10\,n_0$.

Finally, none of these works links its released tables systematically to a posterior.
The five Brussels-extended tables were hand-picked at the extremes of the effective-mass distribution, two of them outside the authors' own refined consistency set \citep{Raduta:2025fpr}, the seven tables of \citet{Du:2021rhq} share a single nuclear-physics parameterization, and the four models of \citet{Alford:2022bpp} are individual fits chosen to span the \ac{xeft} uncertainty of neutron matter.
Our ensemble members are catalog \acp{EOS} of high posterior density, ranked by a deterministic max--min construction into a defined reduction of the multimessenger posterior, and each is delivered with effective-mass variants that vary the thermal response at fixed cold-sector behavior.
The selection step takes only a weighted catalog as input and applies unchanged to any \ac{EOS} posterior, however produced.

At the same time, the \sro{} functional we employ has several shortcomings compared to more modern Skyrme-type functionals.
In particular, the simple \sro{} form of the effective mass (\cref{eq:meff}) cannot reproduce the non-monotonic density dependence that \ac{xeft} predicts in neutron matter \citep{Keller:2020qhx}, so the thermal sector of our tables, including its effective-mass variants, is not constrained by \ac{xeft}.
The consequences for the bulk merger dynamics are limited, because the cold pressure dominates at supranuclear densities \citep{Jacobi:2023olu,Raithel:2025geu} and the threshold mass for prompt collapse is insensitive to realistic thermal variations \citep{Raithel:2025geu}.
The thermal sector does, however, shape the temperature and compactness of the remnant \citep{Raithel:2021hye,Fields:2023bhs}, shifts the postmerger gravitational-wave frequencies by a few percent \citep{Bauswein:2010dn,Raithel:2023zml,Han:2025pho}, and can change the dynamical ejecta mass \citep{Raithel:2023zml,Raithel:2025geu,Han:2025pho}.
It matters more still for applications that depend strongly on the thermal response, such as core-collapse supernovae \citep{Yasin:2018ckc,Schneider:2019shi}.
The variants are intended for systematic studies of exactly these effects rather than as an uncertainty estimate.
Furthermore, nucleonic energy functionals with better-behaved high-density limits exist: the Skyrme-like extensions of \citet{Huth:2020ozf} are built to approach a causal asymptote (a property relativistic mean-field models share by construction), and related forms underpin the energy-density-functional inference of \citet{Lim:2018bkq}, the nucleonic metamodel \citep{Margueron:2017eqc}, and the Skyrme inference with a metamodel high-density extension of \citet{Klausner:2026foh}.
The functional of \citet{Huth:2020ozf} has recently been propagated into finite-temperature general-purpose tables for core-collapse supernova simulations \citep{Jost:2026wcd}.
With a table generator built on a functional of this class, the causality search would simplify and the thermal sector could be made consistent with \ac{xeft} at once, and we consider this the natural next step for constraint-consistent table sets.

The smooth Skyrme functional adopted here describes purely nucleonic matter with a quadratic isospin dependence, and by construction represents neither first-order phase transitions nor non-nucleonic degrees of freedom (hyperons, deconfined quark matter).
The prior, and hence every posterior statement in this work, is conditional on this model space, and the quoted credible intervals do not bound the softening a strong hybrid transition could introduce.
Because hybrid and phase-transition constructions are conventionally built by grafting a high-density branch onto a baryonic reference \ac{EOS}, the released ensemble can serve as such a reference set and be amended with these extensions after the fact, while a full Bayesian treatment of that larger model space is left to future work.

Future measurements can be included as additional or updated likelihood weights on the existing catalog, and the selection can then be repeated without new sampling.
We expect the ensemble and the selection method to provide a foundation for future simulation campaigns of neutron-star mergers and core-collapse supernovae.

\begin{acknowledgments}

MJ and SB acknowledge funding from the Deutsche Forschungsgemeinschaft (DFG) project MEMI number BE 6301/2-2 (project number 443239082).
GH and SB acknowledge funding from the EU Horizon under ERC Consolidator Grant, no.\ InspiReM-101043372 and from the Deutsche Forschungsgemeinschaft, DFG, project MEMI number BE 6301/2-1.
This work was supported by the Deutsche Forschungsgemeinschaft (DFG) - BE 6301/8-1 (project number 563106943) and the National Science Foundation (NSF) - PHY-2512802 under the DFG-NSF Physics ``Multimessenger Astronomy of Neutron Star
Mergers with Numerical Relativity''.
SB acknowledges support from DFG BE 6301/8-1 (project number 563106943).
DR acknowledges support from the U.S.~Department of Energy, Office of Science, Division of Nuclear Physics under Award Number(s) DE-SC0024388, and from the National Science Foundation under Grants PHY-2020275, PHY-2116686, PHY-2407681, PHY-2512802, and PHY-2621752.

Simulations were performed on the Lichtenberg II cluster at TU Darmstadt and the ARA and DRACO clusters at Friedrich Schiller University Jena.
The authors gratefully acknowledge the computing time provided to them at the NHR Center NHR4CES at TU Darmstadt (project number {\tt p0026834}).
This is funded by the Federal Ministry of Research, Technology and Space, and the state governments participating on the basis of the resolutions of the GWK for national high performance computing at universities (\url{www.nhr-verein.de/unsere-partner}).
The ARA cluster is funded in part by DFG grants INST 275/334-1 FUGG and INST 275/363-1 FUGG, and ERC Starting Grant, grant agreement no.\ BinGraSp-714626.

The authors used the AI assistant Claude (Anthropic, several versions of the Claude Opus and Claude Fable models) for code development, figure preparation, and the drafting and revision of the text.
They directed this work, made all modeling and analysis decisions, tested the code and validated every result entering this article, reviewed and edited the full text, and take full responsibility for its content.
\end{acknowledgments}

\section*{Data availability}

The ranked selection list of \cref{tab:selection} and the 36 \sro{} general-purpose tables, one per member and effective-mass level, will be released on Zenodo upon publication of this article.
The modified version of the \sro{} code used to generate the tables is available at \url{https://github.com/Max-Jacobi/sroeos}, with the changes documented in its README.
The full \ac{EOS} catalog and the analysis scripts are available from the authors upon reasonable request.

\appendix

\section{Skyrme coefficient linear system}
\label{app:linear_system}

Given fixed density exponents $\{\delta_i\}$ ($N_{\rm terms} = 8$ for the five-anchor scheme) and target values for the $2N_{\rm terms}$ nuclear-physics observables in \cref{tab:ranges}, the Skyrme coefficients $(a_i, b_i)$ are determined by solving two decoupled $N_{\rm terms}\times N_{\rm terms}$ linear systems, one per isospin channel.

\paragraph{SNM channel.}
Using $\tilde{a}_i = a_i + b_i$, the energy at $x = 1/2$ is $\veps_{\rm pot}^{\rm SNM} = \sum_i \tilde{a}_i n^{\delta_i}$.
The prescribed targets $(B,\, n_0,\, K,\, \csqs{1},\ldots,\csqs{m})$ yield the linear system $\mathsf{M}_{\rm SNM}\,\tilde{\mathbf{a}} = \mathbf{r}_{\rm SNM}$ with row vectors
\begin{align}
  [\mathsf{M}_{\rm SNM}]_{B}       &= (n_0^{\delta_i})\,, \\
  [\mathsf{M}_{\rm SNM}]_{n_0}     &= (\delta_i\, n_0^{\delta_i})\,, \\
  [\mathsf{M}_{\rm SNM}]_{K}       &= (\delta_i(\delta_i+1)\, n_0^{\delta_i})\,, \\
  [\mathsf{M}_{\rm SNM}]_{\csqs{j}} &= \bigl((\delta_i+1)\, n_{j}^{\delta_i}
                                        (\delta_i - \csqs{j})\bigr)\,,
\end{align}
where $j = 1,\ldots,m$, $n_j$ runs over the $m$ anchor densities of \cref{tab:ranges} and $\csqs{j}$ denotes the sampled \ac{SNM} target $c_s^2(n_j, 1/2)$.
The right-hand side components are
\begin{align}
  r_B      &= B + \Delta/2 - \veps_{\rm kin}(n_0, \tfrac{1}{2})\,, \\
  r_{n_0}  &= -P_{\rm kin}(n_0, \tfrac{1}{2})/n_0\,, \\
  r_K      &= K/9 - (\partial P_{\rm kin}/\partial n)|_{n_0,\, 1/2}\,, \\
  r_{\csqs{j}} &= \csqs{j}\, h_{\rm kin}(n_j, \tfrac{1}{2})
                   - (\partial P_{\rm kin}/\partial n)|_{n_j,\, 1/2}\,,
\end{align}
where all kinetic quantities are evaluated with the effective masses of \cref{eq:meff}, at the $\alpha_{1,2}$ values fixed in closed form by the $m^*$ prescription of \cref{sec:model}.
The last row follows from the causality equation $c_s^2 = (\partial P/\partial n)_x / h$ evaluated at $n = n_j$, $x = 1/2$.
Here $h_{\rm kin}$ follows the same convention as $h$ in \cref{eq:cs2_def} with only kinetic (and electron) contributions retained, $h_{\rm kin} = m_n + \veps_{\rm kin} + P_{\rm kin}/n - x\Delta + x\,\mu_e$.
In particular it includes the neutron rest mass.
Likewise, $\partial P_{\rm kin}/\partial n$ contains the electron term $x\,\partial P_e/\partial n_e$.

\paragraph{PNM channel.}
Only the coefficients $a_i$ appear because at $x = 0$ the interaction reduces to $\veps_{\rm pot}^{\rm PNM} = \sum_i a_i n^{\delta_i}$.
The prescribed targets $(J,\, L,\, \tilde{K}_{\rm sym},\, \csqp{1},\ldots,\csqp{m})$ yield $\mathsf{M}_{\rm PNM}\,\mathbf{a} = \mathbf{r}_{\rm PNM}$ with the same row structure as above (replacing $x = 1/2 \to 0$ and the target values accordingly), except for three right-hand sides: the $J$-row uses $r_J = B + J - \veps_{\rm kin}(n_0, 0)$, the $L$-row (the \ac{PNM} analog of the \ac{SNM} $n_0$-row) uses $r_L = L/3 - P_{\rm kin}(n_0, 0)/n_0$, and the curvature row uses $r_{\tilde{K}_{\rm sym}} = (K + \tilde{K}_{\rm sym})/9 -(\partial P_{\rm kin}/\partial n)|_{n_0,\, 0}$.
The two systems are solved independently by standard $\mathsf{LU}$ decomposition.

\section{Exponent existence search: full protocol}
\label{app:search}

\Cref{sec:search} introduces the SLSQP existence search at a conceptual level.
This appendix gives the numerical protocol used for the production catalog of \cref{sec:results}.

\paragraph{Objective and constraints.}
The descent minimizes
\begin{align}
  \label{eq:search_objective}
  F([\delta_i]) ={}& \frac{1}{G} \sum_{(n,x)} \Bigl[\max\bigl(0,\, m - c_s^2(n,x)\bigr)^{3/2} \nonumber\\
                   &\qquad + \max\bigl(0,\, c_s^2(n,x) - (1-m)\bigr)^{3/2}\Bigr] \nonumber\\
                   &- \lambda \sum_i \ln(\delta_i - \delta_{i-1})\,,
\end{align}
where $m = 0.02$ is a small margin inside the causal interval, $\lambda = 10^{-2}$, and the sum runs over a grid of $G$ points, 55 densities evenly spaced in $n/n_0$ from $0.5$ to $11$ (one $n_0$ beyond the topmost speed-of-sound anchor) times 21 values of $x$ evenly spaced in $[0,0.5]$.
The analytic dependency of $c_s^2(n,x)$ on $[\delta_i]$ is given by the linear solve of \cref{app:linear_system}.
The first term is a hinge penalty of power $1.5$ on $c_s^2$ excursions outside $[m, 1-m]$.
A quadratic hinge (power 2) was found to stall on shallow violations (vanishing gradient as the violation depth $\to 0$).
A linear hinge (power 1) corrupts the optimizer's quasi-Newton model with gradient discontinuities at the hinge. Power $1.5$ keeps the objective continuously differentiable while remaining sensitive to shallow violations.
The second term is a smooth log-barrier on the consecutive exponent gaps (including the fixed $\delta_1 = 1$), which steers the optimizer toward well-separated exponents without rejecting draws, backed by a hard minimum-gap constraint of $0.02$ per pair.
The hard constraint is active for part of the catalog. Two of the twelve delivered members have exponent pairs at the minimum gap, among them the member ranked sixth, whose coefficients reach $5\times10^{12}$ and whose round-off error is described in \cref{sec:selection}. We therefore release all coefficients and exponents at full double precision.
It counters near-degenerate exponent pairs, which give large, canceling Skyrme coefficients that ill-condition the linear solve of \cref{app:linear_system}.
The barrier lowers the surviving catalog's linear-solve condition numbers by over an order of magnitude in the bulk at unchanged survivor yield and with no measurable shift of the effective prior.
Small positive margins ($0.1$\,MeV each) on the unitary-gas and symmetry-energy constraints of \cref{sec:search} ensure that the optimizer's own convergence tolerance cannot leave a solution just outside the bound.
The coupling bound $|v_{a,b}| \le 500\,{\rm MeV\,fm^3}$ is imposed two-sided on 48 logarithmically spaced densities in $[10^{-4}, 0.05]\,{\rm fm^{-3}}$ and is tested again at acceptance, so that a soft failure of the optimizer cannot leak through.
At acceptance, the pure hinge term is tested on a finer grid (240 density points to $10\,n_0$, the topmost anchor) than the objective above (which extends one $n_0$ further, nudging solutions to stay causal a little past their last explicit constraint).

\paragraph{Restarts and starting points.}
Up to 5 independent restarts are attempted per draw, stopping at the first one that finds a causal, constraint-satisfying solution.
Each restart draws a fresh set of starting exponent gaps ($\delta_i - \delta_{i-1}$) from an exponential distribution and screens up to 20 such starting points before committing to a full descent.
The screen accepts a starting point if $\max |c_s^2 - 1/2| < 1.5$ over the objective grid, i.e.\ if $c_s^2 \in (-1, 2)$ everywhere, at the cost of a single objective evaluation per guess.
Starting points failing this test lie so far from the causal band that descents from them rarely succeed.
If no guess passes within the budget, the least-bad one (smallest $\max |c_s^2 - 1/2|$) is used instead of vetoing the draw.
The selected starting point then seeds a single SLSQP descent (20 iterations, analytic Jacobian) run to completion.

\paragraph{Polish stage.}
If every restart fails but the best attempt is a near-miss (final penalty below a small threshold), one basinhopping chain (repeated perturb-and-redescend hops, each a further local SLSQP polish) is seeded at that near-miss before giving up on the draw.

\paragraph{Coupling refinement.}
Every accepted exponent vector seeds one further SLSQP descent (up to 100 iterations) on the search objective augmented by a log-hinge penalty on $\max |v_{a,b}|$ with its knee at $300\,{\rm MeV\,fm^3}$, under the same hard constraints.
The refined vector replaces the accepted one only if it stays causal on the acceptance grid and lowers $\max |v_{a,b}|$, so this pass cannot change which draws are accepted and only moves accepted solutions deeper inside the bound.
It was added because some accepted members close to the ceiling still failed table generation, and the refinement recovers about half of those.

\section{BPS crust matching}
\label{app:bps}

The matching procedure operates in the space of total energy density $e$ (rest mass included) and pressure $P$.
First, we find $e_c$, defined by minimizing the pressure difference $|P_{\rm core}(e_c ) - P_{\rm BPS}(e_c)|$ over the overlap in energy density.
The corresponding baryon density $n_c$ is then obtained by linear interpolation.
Between $n_c - \Delta n$ and $n_c + \Delta n$ we connect $P_{\rm core}(e)$ and $P_{\rm BPS}(e)$ with a Piecewise Cubic Hermite Interpolating Polynomial (PCHIP) spline fitted to both the BPS data below $n_c - \Delta n$ and the Skyrme-core data above $n_c + \Delta n$.
Starting from $\Delta n = 0.05\,$fm$^{-3}$, the gap is widened iteratively by a factor $1.5$ each step until the pressure increase across the gap, $\Delta P = P(n_c + \Delta n) - P(n_c - \Delta n)$, exceeds a minimum threshold of $0.01$\,MeV\,fm$^{-3}$.
The resulting composite \ac{EOS} is smooth and monotone across the transition by construction.
This matched barotrope $P(e)$ is all the catalog stage needs, since it enters only the cold \ac{TOV} solve.
The released finite-temperature tables do not use it. There, the low-density inhomogeneous phase and all its thermodynamic derivatives are computed within the \sro{} code itself (\cref{sec:selection}).

\section{Astrophysical importance estimator}
\label{app:estimator}

The GW and NICER channels of \cref{sec:astro} share one estimator.
For each event $k$ we fit a Gaussian \ac{KDE} $\hat p_k$ to the published posterior samples in the event's own measurement space: $(M, R)$ for the NICER sources, and chirp mass, mass ratio, and the two tidal deformabilities for GW170817.
The \ac{KDE} is evaluated at the point the candidate \ac{EOS} predicts for given source masses: the radius $R(m; i_{\rm EOS})$ read off the candidate's $M$--$R$ curve, or the deformabilities $\Lambda_{1,2}(m_{1,2}; i_{\rm EOS})$ off its $\Lambda(M)$ curve.
The source masses are nuisance parameters and are marginalized with a flat prior, reusing the event's own posterior mass samples $m^{(s)}$ as importance points:
\begin{align}
  \label{eq:astro_L}
  p(\mathbf{d}_k\,|\,i_{\rm EOS}) &\propto
  \frac{\sum_s w_s\, \hat p_k\bigl(\mathbf{x}(m^{(s)};\, i_{\rm EOS})\bigr)}
       {\sum_s w_s}\,,
  \\
  w_s &= \frac{1}{\hat p^{\,m}_k(m^{(s)})}\,,
  \nonumber
\end{align}
where $\mathbf{x}(m; i_{\rm EOS})$ is the predicted measurement-space point and $\hat p^{\,m}_k$ is a \ac{KDE} of the event's marginal mass distribution.
The weights $w_s$ cancel the sampling density of the mass samples, so the mass integral runs over a flat prior rather than over the event's own mass posterior: a plain average over the samples would implicitly re-impose that posterior as a mass prior and double-count the mass information.
For GW170817, $w_s$ additionally contains the Jacobian from the sampled $(\mathcal{M}, q)$ coordinates to component masses, keeping a flat prior in $(m_1, m_2)$.
Since $\hat p_k$ is built from posterior samples, it estimates the event's posterior, not its likelihood.
Using it in place of the likelihood is justified if the original analysis' prior is flat in the remaining measurement coordinates (radius for the NICER sources, tidal deformabilities for GW170817), so that posterior and likelihood are proportional there.
The published priors for these data sets are approximately flat in those coordinates.

\section{Repair of the single-nucleus solve}
\label{app:sna_repair}

The \ac{SNA} solve of \sro{} that underlies the finite-temperature tables (\cref{sec:selection}) only accepts heavy nuclei with $Z > 6$, which ends the inhomogeneous phase while nuclei are still more bound than uniform matter.
In isolated groups of cells near this boundary the solver also converges to a root that is not the one of lowest free energy.
Both leave steps in pressure and internal energy at the crust-core transition, and the tables distributed with the \sro{} code show the same pressure decreasing with density across the boundary.

We therefore solve the cells at the boundary a second time, starting from the first-pass solution of neighboring cells, accepting nuclei down to $Z > 3$, and keeping the solution with the lower free energy.
At temperatures $T \lesssim 1$\,MeV the inhomogeneous solve fails in a single density column just past the cold edge of the nuclear region, although the same column holds nuclei at higher temperature.
We fill these cells, at most $0.06\%$ of the \ac{SNA} grid, from the nearest proton-fraction row that is nuclear at the same density, continuing the free energy with that row's temperature dependence and deriving the entropy, the energy and the temperature derivatives from it, so that the filled cells are thermodynamically consistent.
For the member ranked third, whose cold crust reaches very low proton fractions, we additionally solve a one-cell layer at the cold edge with the charge limit relaxed to $Z > 0$.

\sro{} obtains the second derivatives of the free energy by implicit differentiation of the equilibrium equations, which become singular at the high-temperature end of the nuclear branch.
We rebuild the derivatives of the affected cells, all within about two temperature rows of that end, from centered finite differences of their neighbors and the Maxwell relations.
In the member ranked sixth, the stiffest in its Skyrme coefficients, the entropy of cold dense uniform matter carries a round-off error of about $0.01\,k_B$ per baryon, which enters the sound speed through $\partial s/\partial T$.
There we replace the affected sound-speed values, less than $0.01\%$ of the table, by interpolation along temperature.

The remaining defects lie at the melting edge of the crust between $1$ and $8$\,MeV, where the $Z > 3$ limit still ends the nuclei before uniform matter is favored.
There the pressure decreases with density in at most $0.01\%$ of the cells of a table, by up to $22\%$ between neighboring density points.
No table has a cell where the energy decreases with temperature, and the sound speed lies between zero and the speed of light everywhere.

\input{paper20261118.bbl}
\end{document}

%% file: selection_table.tex
%% generated by scripts/selection_table.py -- do not edit
1 & 185 & 34.3 & 46.9 & $-$204 & 0.45 & 0.66 & 0.051 & 11.74 & 317 & 2.10 & 11.02 & 11.37 & 1.00 \\
2 & 293 & 31.1 & 38.6 & $-$362 & 0.46 & 0.27 & 0.024 & 11.70 & 342 & 2.16 & 11.74 & 10.30 & 7.99 \\
3 & 227 & 31.7 & 38.6 & $-$271 & 0.37 & 0.82 & 0.025 & 11.30 & 261 & 2.29 & 10.76 & 10.45 & 2.25 \\
4 & 298 & 33.2 & 52.9 & $-$246 & 0.51 & 0.86 & 0.059 & 12.21 & 424 & 2.40 & 11.35 & 10.86 & 1.13 \\
5 & 219 & 32.3 & 51.6 & $-$81 & 0.49 & 0.42 & 0.066 & 12.29 & 437 & 2.04 & 11.66 & 10.51 & 0.73 \\
6 & 193 & 35.0 & 64.0 & $-$228 & 0.35 & 0.90 & 0.061 & 11.42 & 247 & 2.24 & 10.49 & 11.15 & 0.25 \\
7 & 301 & 33.8 & 40.8 & $-$397 & 0.36 & 0.58 & 0.028 & 11.13 & 239 & 2.04 & 10.78 & 11.58 & 1.67 \\
8 & 224 & 34.4 & 56.1 & $-$209 & 0.36 & 0.81 & 0.064 & 11.37 & 239 & 2.02 & 10.24 & 11.83 & 0.84 \\
9 & 267 & 32.5 & 35.8 & $-$229 & 0.51 & 0.38 & 0.043 & 12.01 & 401 & 2.11 & 11.69 & 11.77 & 1.50 \\
10 & 224 & 34.5 & 65.9 & $-$41 & 0.43 & 0.73 & 0.081 & 12.31 & 407 & 2.10 & 11.10 & 11.69 & 0.21 \\
11 & 269 & 31.4 & 32.5 & $-$339 & 0.48 & 0.61 & 0.029 & 11.91 & 386 & 2.22 & 11.54 & 10.64 & 0.82 \\
12 & 302 & 30.5 & 47.8 & $-$126 & 0.54 & 0.67 & 0.060 & 12.85 & 609 & 2.32 & 12.00 & 10.35 & 0.0088